\documentclass[a4paper,twocolumn,superscriptaddress,11pt,accepted=2017-06-13]{quantumarticle}
\pdfoutput=1
\usepackage[utf8]{inputenc}
\usepackage[english]{babel}
\usepackage[T1]{fontenc}
\usepackage{amsmath, amsthm, amssymb, bbm, amsfonts, graphicx, braket, framed, changes}
\usepackage{amsmath}
\usepackage{hyperref}
\usepackage{tikz}

\newcommand{\unity}{\ensuremath{\mathbbm{1}}}
\def\pud{\mathrm{PU} (d)}
\def\pcd{\mathrm P \mathbb C ^d}
\def\F{\mathcal F}
\def\be{\mathbf{e}}
\def\ud{\mathrm{U}(d)}
\def\sud{\mathrm{SU}(d)}

\theoremstyle{definition}
\newtheorem{result}{Result}

\newtheorem{lemma}{Lemma}

\usepackage[numbers,sort&compress]{natbib}

\begin{document}

\title{Classification of all alternatives to the Born rule 
in terms of informational properties}
\date{\today}
\author{Thomas D. Galley}
\email{thomas.galley.14@ucl.ac.uk}
\affiliation{Department of Physics and Astronomy, University College London,
Gower Street, London WC1E 6BT, United Kingdom}
\orcid{0000-0002-8870-3215}
\author{Lluis Masanes}
\affiliation{Department of Physics and Astronomy, University College London,
Gower Street, London WC1E 6BT, United Kingdom}
\orcid{0000-0002-1476-2327}

\maketitle

\begin{abstract}
The standard postulates of quantum theory can be divided into two groups: 
the first one characterizes the structure and dynamics of pure states, while the second one specifies the structure of measurements and the corresponding probabilities.
In this work we keep the first group of postulates and characterize all alternatives to the second group that give rise to finite-dimensional sets of mixed states. We prove a correspondence between all these alternatives and a class of representations of the unitary group. 
Some features of these probabilistic theories are identical to quantum theory, but there are important differences in others. 
For example, some theories have three perfectly distinguishable states in a two-dimensional Hilbert space.
Others have exotic properties such as lack of ``bit symmetry'', the violation of ``no simultaneous encoding'' (a property similar to information causality) and the existence of maximal measurements without phase groups. We also analyze which of these properties single out the Born rule.
\end{abstract}

\section{Introduction}

There is a long-standing debate on whether the structure and dynamics of pure states already encodes the structure of measurements and probabilities. 
This discussion arises within the dynamical description of a quantum measurement, decoherence theory, and the many-worlds interpretation.
There have been attempts to derive the Born rule~\cite{Deutsch_quantum_1999, Zurek_probabilities_2005, Wallace_how_2010}, however these are often deemed controversial. In this work we suggest a more neutral approach where we consider \emph{all} possible alternatives to the Born rule and explore their physical and informational properties.

More precisely, we provide a complete classification of all the alternatives to the structure of measurements and the formula for outcome probabilities with the following property: in any system with a finite-dimensional Hilbert space, the number of parameters that is required to specify a mixed state is finite.
This property is necessary if state estimation can be performed with a finite number of measurements and no additional assumptions.
We also show that, among all these  alternative theories, those which have no restriction on the allowed effects violate a physical principle called ``bit symmetry''~\cite{Muller_power_2011}, which is satisfied by quantum theory. Bit symmetry states that any pair of perfectly distinguishable pure states can be reversibly mapped to any other. Thus the Born rule is the only possible probability assignment to measurement outcomes under these requirements. We hope that these results may help to settle the above mentioned debate. 

Most other known non-quantum general probabilistic theories seem somewhat pathological~\cite{Gross_all_2010,Sabri_reversible_2014, Sabri_reversibility_2015} with some exceptions such as quantum theory over the field of real numbers~\cite{Hardy_limited_2012} or quaternions~\cite{finkelstein_notes_1959} and theories based on euclidean Jordan algebras~\cite{Barnum_composites_2016}.

For example they often violate most of the axioms used to reconstruct quantum theory, rather than just one or two per reconstruction.
In particular, modifying the Born rule without creating inconsistencies or absurdities has been argued to be difficult (see, e.g.~\cite{Aaronson_quantum_2004}). 

For example, if we keep the usual association of the outcomes of a measurement with the elements of an orthonormal basis $\{ |k\rangle \}$, a straightforward alternative for the probability of $|k\rangle$ given sate $|\psi\rangle $ is
\begin{equation}\label{alternative 1}
  P(k|\psi) = 
  \frac {|\langle k| \psi\rangle|^\alpha} {\sum_{k'} |\langle k'| \psi\rangle|^\alpha}\ ,
\end{equation}
with $\alpha\neq 2$. However, this violates the finiteness assumption stated above.

In this work, we provide many non-quantum probabilistic theories with a modified Born rule which still inherit many of quantum theory's properties. 
These theories can be used as foils to understand quantum theory ``from the outside''.
In particular, we analyse in full detail all finite-dimensional state spaces where pure states are rays in a 2-dimensional Hilbert space. 
We discuss properties such as the number of distinguishable states, bit symmetry, no-simultaneous encoding and phase groups for maximal measurements. 
We also analyze a class of alternative theories which have a restriction on the allowed effects  such that they share all the above properties with quantum theory.

In section~\ref{sectiontheorem} we introduce the necessary framework and outline the main result showing the equivalence between alternative sets of measurement postulates and certain representations of the dynamical group. We then find these representations. In section~\ref{sectionphenomenology} we study these alternative theories starting with a classification of theories with pure states given by rays on a 2-dimensional Hilbert space. We then show that all unrestricted non-quantum theories with pure states given by rays on a $d$-dimensional Hilbert space violate bit symmetry. In section~\ref{sectionconclusion} we discuss the results and their relation to Gleason's theorem. Proofs of all the results are in the appendix.

\section{All alternatives to the Measurement Postulates}\label{sectiontheorem}

\subsection{Outcome probability functions}

The standard formulation of quantum theory can be divided in two parts. The first part postulates that the pure states of a system are the rays of a complex Hilbert space\footnote{The set of rays of a Hilbert space $\mathcal H$ is called the projective Hilbert space, denoted $\mathrm P \mathcal H$.} $\psi \in \pcd$, and that the evolution of an isolated quantum system is unitary $\psi \mapsto  U \psi$, where $U\in \pud$ is an element of the projective unitary group\footnote{$\pud$ is  defined by taking $\mathrm{SU}(d)$ and identifying the matrices that differ by a global phase $U\cong e^{i \theta} U$. For example, the two matrices $\pm \unity \in \mathrm{SU}(2)$ correspond to the same element in $\mathrm{PU}(2)$.}. 
The second part postulates that each measurement is characterized by a list of positive semi-definite operators $(Q_1, Q_2, \ldots, Q_k)$ adding up the identity $\sum_i Q_i = \unity$, and that the probability of outcome $Q_i$ when the system is in state $\psi$ is given by $P(Q_i|\psi) = \langle \psi | Q_i |\psi\rangle$ (Born's rule).

In this article we consider \emph{all} possible alternatives to this second part. For this, we denote by $P(F|\psi)$ the probability of outcome $F$ given the pure state $\psi$.
And we recall that, from an operational perspective, the mathematical characterization of an outcome $F$ is given by the probability of its occurrence for all states. Hence, each outcome $F$ is represented by the function $F: \pcd \to \mathbb [0,1]$ defined as $F(\psi) = P(F|\psi)$.
Note that, unlike in~\cite{Han_Quantum_2016}, we do not require outcomes $F$ to correspond to those of quantum mechanics (POVM elements).
And even more, a priori, these outcome probability functions (OPFs) $F$ are arbitrary; for example, they can be non-linear in $|\psi\rangle\! \langle\psi|$ or even discontinuous. 
However, we show below that, if mixed states are required to have finitely-many parameters then OPFs must be polynomial in $|\psi\rangle\! \langle\psi|$.

In this generalized setup, each measurement is characterized by a list of OPFs $(F_1, F_2, \ldots, F_k)$ satisfying the normalization condition $\sum_i F_i (\psi) = 1$ for all $\psi \in \pcd$.
Therefore, each alternative to the Measurement Postulates can be characterized by the list of all ``allowed'' measurements in each dimension $d$, e.g. $\{(F_1, \ldots, F_k),(F'_1,\ldots, F'_{k'}),\ldots \}$. 
For each alternative, we denote the set of OPFs on $\pcd$ by $\F _d = \{F_1, F_2,\ldots, F'_1, F'_2, \ldots \}$.
From an operational perspective, the set $\F_d$ is not completely arbitrary: the composition of a measurement and a unitary generates another measurement. That is, if $F \in \F_d$ and $U\in \pud$ then $F\circ U \in \F_d$. 

In this generalized setup, a \emph{mixed state} is an equivalence class of indistinguishable ensembles. Let $(p_i, \psi_i)$ be the ensemble where state $\psi_i \in\pcd$ is prepared with probability $p_i$. Two ensembles $(p_i, \psi_i)$ and $(p'_i, \psi'_i)$ are indistinguishable in $\mathcal F_d$ if
\begin{equation}
  \sum_i p_i F(\psi_i)
  =
  \sum_i p'_i F(\psi'_i)
\end{equation}
for all $F\in \mathcal F_d$. This equation defines the equivalence relation with which the equivalence classes of ensembles (mixed states) are defined.

\subsection{The convex representation}

In what follows we introduce a different representation for pure states, unitaries and OPFs, which naturally incorporates mixed states.
This representation is the one used in general probabilistic theories~\cite{Hardy_quantum_2001, Barrett_information_2005, Masanes_derivation_2011,Dakic_quantum_2009}, which is sometimes called 
the convex operational framework~\cite{giles_foundations_1970, mielnik_generalized_1974, Chiribella_probabilistic_2010, Barnum_entropy_2010}.
Before starting let us introduce some notation. Given a real vector space $V$, the general linear group $\mathrm{GL} (V)$ is the set of invertible linear maps $T: V\to V$, and $\mathcal{E} (V)$ is the set of affine\footnote{A function $E :V \to \mathbb R$ is affine if it satisfies $E \left( x \omega_1 +(1-x) \omega_2 \right) = x E(\omega_1) + (1-x) E(\omega_2)$ for all $x\in \mathbb R$ and $\omega_1, \omega_2 \in V$. When $V$ is finite-dimensional, each $E \in \mathcal{E} (V)$ can be written in terms of a scalar product as $E(\omega) = \be\cdot \omega +c$, where $\be \in V$ and $c\in \mathbb R$.} functions $E: V\to \mathbb R$.

\begin{figure}
     \begin{center}
     	\includegraphics[width = 0.6 \columnwidth]{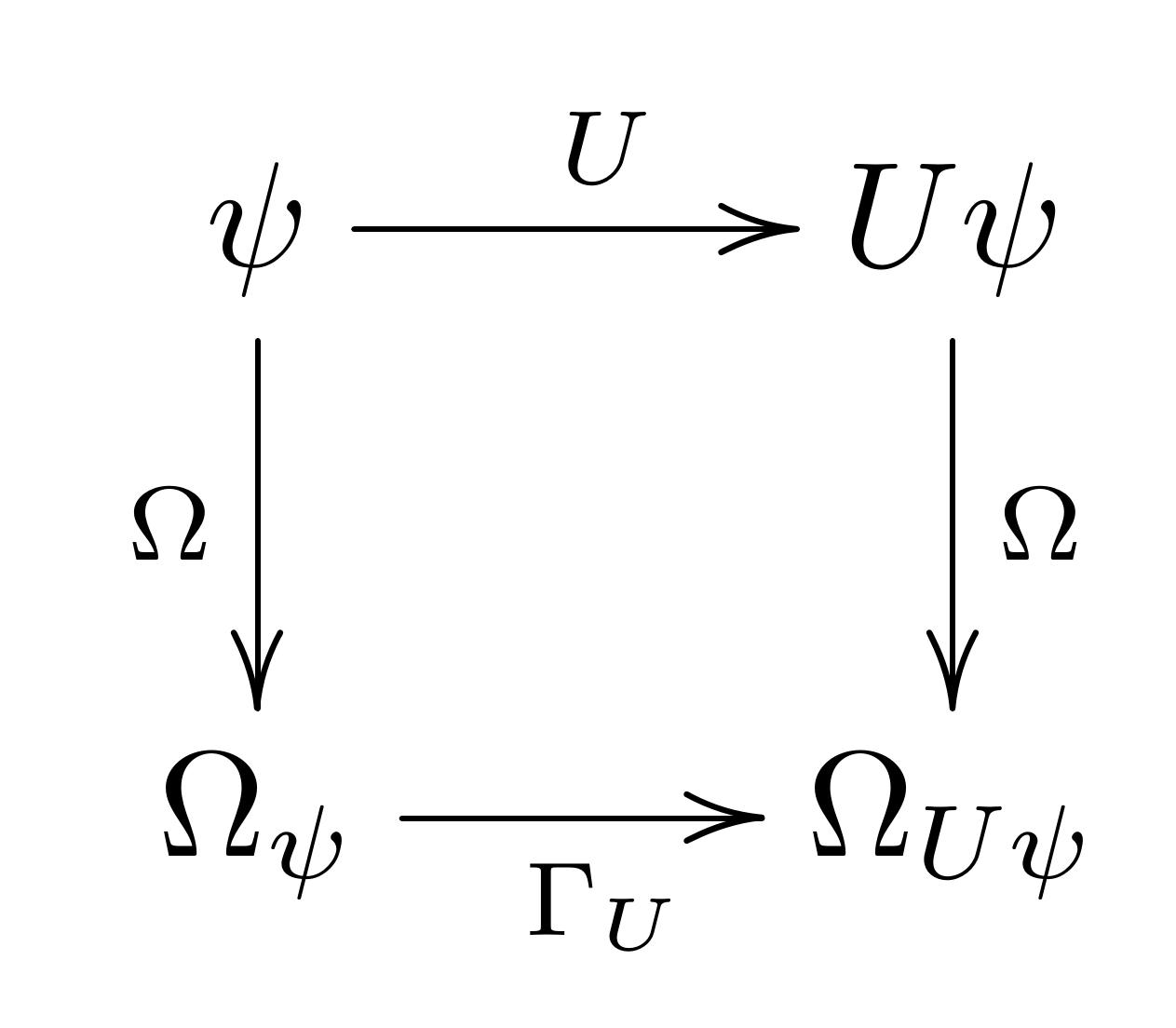}
     	\caption{This diagram expresses the commutation of equation~\eqref{dynam struc 2}.	
     }
     	\label{ComFig}
     \end{center}
\end{figure}

\begin{result}\label{R1}
Given a set $\F_d$ of OPFs for $\mathrm P \mathbb C^d$ (encoding an alternative to the measurement postulates) there is a (possibly infinite-dimensional) real vector space $V$ and the maps
\begin{align}
  \Omega &: \pcd \to V\, ,
  \\
  \Gamma &: \pud \to \mathrm{GL} (V)\, ,
  \\
  \Lambda &: \F_d \to \mathcal{E}(V)\, ,
\end{align}
satisfying the following properties:
\begin{itemize}
  \item \emph{Preservation of dynamical structure (see Figure~\ref{ComFig}):}
\begin{align}
  \label{dynam struc 2}
  \Gamma_U \Omega_\psi &= \Omega_{U\psi}\, ,
  \\ \label{dynam struc 1}
  \Gamma_{U_1} \Gamma_{U_2} &= \Gamma_{U_1 U_2}\, .
\end{align}

  \item \emph{Preservation of probabilistic structure:}
\begin{equation}
  \Lambda_F \! \left( \Omega_\psi \right)
  =
  F(\psi)
  \, . 
\end{equation}

  \item \emph{Minimality of $V$\footnote{Here Aff($S$) refers to the affine span of vectors in $S$. The affine span is defined as all linear combinations of vectors in $S$ where the coefficients add to 1.}:}
\begin{equation}
  \label{span O}
  {\rm Aff}\, (\Omega_{\pcd})
  = V\ .
\end{equation}

  \item \emph{Uniqueness:} for any other maps $\Omega', \Gamma', \Lambda'$ satisfying all of the above, there is an invertible linear map $L:V\to V$ such that
\begin{align}
  \label{uni1}
  \Omega'_\psi &= L  (\Omega_\psi) \, ,
  \\   \label{uni2}
  \Gamma'_U &= L   \Gamma_U   L^{-1} ,
  \\   \label{uni3}
  \Lambda'_F &= \Lambda_F   L^{-1} .
\end{align}
\end{itemize}
\end{result}

\noindent
In this new representation, the affinity of the OPFs $\Lambda_F$ implies the following. All the predictions for an ensemble $(p_i, \psi_i)$ can be computed from the vector $\omega = \sum_i p_i \Omega_{\psi_i} \in V$. In more detail, if a source prepares state $\psi_i$ with probability $p_i$ then the probability of outcome $F$ is
\begin{align}
  \nonumber
  \sum_i p_i P(F|\psi_i) 
  = 
  \sum_i p_i \Lambda_F (\Omega_{\psi_i}) 
  =  
  \Lambda_F( \omega)
  \ .
\end{align}
Hence, the ``mixed state'' $\omega$ contains all the physical information of the ensemble, in the same sense as the density matrix $\rho = \sum_i p_i |\psi_i\rangle\! \langle\psi_i |$ does in quantum theory.
More precisely, $\omega$ is the analogue of the traceless part of the density matrix $\rho - \unity \frac {{\rm tr} \rho} d$, which in the $d=2$ case is the Bloch vector. 
So $V$ is the analogue of the space of traceless Hermitian matrices or Bloch vectors, not the full space of Hermitian matrices.
This is why probabilities are affine instead of linear functions.
For example,in this representation, the maximally mixed state is always the zero vector $\int\! d \psi\, \Omega_\psi = 0 \in V$.
Also, the linearity of the action $\Gamma_U:V\to V$ implies that, transforming the ensemble as $(p_i, \psi_i) \to (p_i, U\psi_i)$, is equivalent to transforming the mixed state as
\begin{align}
  \nonumber
  \sum_i p_i \Omega_{U\psi_i} 
  = 
  \sum_i p_i \Gamma_U \Omega_{\psi_i}
  =  
  \Gamma_U \omega 
  \ .
\end{align}
When we include all these mixed states the (full) state space $\mathcal S$ becomes convex:
\begin{equation}
  \mathcal S = 
  {\rm conv}\, \Omega_{\pcd} 
  \subset V\ .
\end{equation}
An \emph{effect} on $\mathcal S$ is an affine function $E: \mathcal S \to [0,1]$. 
In the representation introduced in Result~\ref{R1}, OPFs $\Lambda_F$ are effects. We say that a given set of OPFs $\F_d$ is \emph{unrestricted}~\cite{Janotta_gen_2013} if it contains all effects of the corresponding state space $\mathcal S$. That is, for every effect $E$ on $\mathcal S$ there is an OPF $F\in \F_d$ such that $\Lambda_F= E$. 
When $\F_d$ is unrestricted the map $\Lambda$ is redundant, and the theory is characterized by $\Omega$ and $\Gamma$, which provide the state space $\mathcal S$.
When $\F_d$ is not unrestricted, the theory can be understood as an unrestricted one with an extra restriction on the allowed effects. Hence, unrestricted theories play a special role.

The dimension of $\mathcal S$ (equal to that of $V$) corresponds to the number of parameters that are necessary to specify a general mixed state.
In quantum theory this number equal to $d^2-1$. 
Despite the finite dimensionality of $\pcd$ the  state space $\mathcal S = {\rm conv}\, \Omega_{\pcd}$ corresponding to a general $\mathcal F_d$, can be infinite-dimensional.
When this is the case, general state estimation requires collecting an infinite amount of experimental data, which makes this task impossible. 
Recall that, when performing state estimation in infinite-dimensional quantum mechanics, additional assumptions are required; like, for example, an upper bound on the energy of the state.
For this reason, in the rest of the paper, we assume that when the set of pure states $\pcd$ is finite-dimensional then so is the set of mixed states $\mathcal S$.

Let us also assume that any mathematical transformation $T:\mathcal S\to \mathcal S$ which can be arbitrarily-well approximated by physical transformations $\Gamma_{\pud}$ should be considered a physical transformation $T\in \Gamma_{\pud}$. This means that the group of matrices $\Gamma_{\pud}$ is topologically closed. Together with the homomorphism property~\eqref{dynam struc 1} and the finite-dimensionality of $V$ this implies that $\Gamma: \pud \to \mathcal \mathrm{GL}(V)$ is a \emph{continuous} group homomorphism (this is shown in appendix \ref{continuity Gamma}). In other words, $\Gamma$ is a representation of the Lie group $\pud$.

\subsection{Classification results}

In quantum mechanics all pure states $\psi$ can be related to a fixed state $\psi_0$ via a unitary $\psi= U\psi_0$. This and equation~\eqref{dynam struc 2} imply that, given the representation $\Gamma$ and the image of the fixed pure state $\psi_0 \to \Omega_{\psi_0}$, we can obtain the image of any pure state as
\begin{equation}
  \Omega_\psi = \Omega_{U\psi_0}
  = \Gamma_U \Omega_{\psi_0}
  \ .
\end{equation}
Now note that, for any OPF $F$, we can write
\begin{equation}
  F(\psi) = F(U\psi_0) = 
  \Lambda_F (\Gamma_U \Omega_{\psi_0})\ .
\end{equation}
This, the affinity of $\Lambda_F$ and the continuity of $\Gamma$ imply that $F: \pcd \to [0,1]$ is continuous.

We also see that, an unrestricted theory is characterized by a representation $\Gamma$ and a reference vector $\Omega_{\psi_0} \in V$.
But not all pairs $(\Gamma, \Omega_{\psi_0})$ satisfy~\eqref{dynam struc 2}. Below we prove that only a small class of representations $\Gamma$ allow for~\eqref{dynam struc 2}, and for each of these, there is a unique (in the sense of~\eqref{uni1}-\eqref{uni3}) vector $\Omega_{\psi_0}$ satisfying~\eqref{dynam struc 2}.

Any representation of $\pud$ is also a representation of $\mathrm{SU}(d)$, and these are all well classified in the finite-dimensional case.
We now consider a family of irreducible representations of $\sud$ which we call $\mathcal D_j^d$. There are many other representations of $\sud$ but these are not compatible with the dynamical structure of quantum theory~\eqref{dynam struc 2}.
For any positive integer $j$, let $\mathcal D_j^d : \mathrm{SU}(d) \to {\rm GL}(\mathbb R^{D_j^d})$ be the highest-dimensional irreducible representation inside the reducible one $\mathrm{Sym}^j U \otimes \mathrm{Sym}^j U^*$, where $\mathrm{Sym}^j U$ is the projection of $U^{\otimes j}$ into the symmetric subspace~\cite[Appendix 2]{Fulton91}. Here $U$ is the fundamental representation of $\mathrm{SU}(d)$ acting on $\mathbb{C}^d$. Note that any global phase $e^{i \theta} U$ disappears in the product $\mathrm{Sym}^j U \otimes \mathrm{Sym}^j U^*$, hence $\mathcal D_j^d$ is also an irreducible representation of $\pud$. 
We recall that these irreducible representations are real, in the sense that there exists a basis in which all matrix elements of $\mathcal D_j^d (U)$ are real, for all $U$.
The dimension $D_j^d$ of the real vector space acted upon by $\mathcal D_j^d (U)$ is 
\begin{equation}\label{dimension}
  D_j^d = \left(\frac{2j}{d-1}+1\right) \prod_{k=1}^{d-2} 
  \left(1 +\frac j k \right)^2
  \ ,
\end{equation}
(see \cite[p.224]{Fulton91}). 

\begin{figure}
\raggedright (a)
\begin{center}
\includegraphics{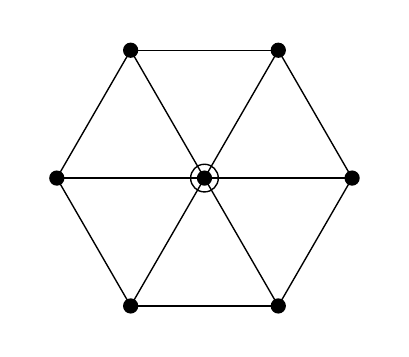}
\end{center}
(b)
\begin{center}
\includegraphics{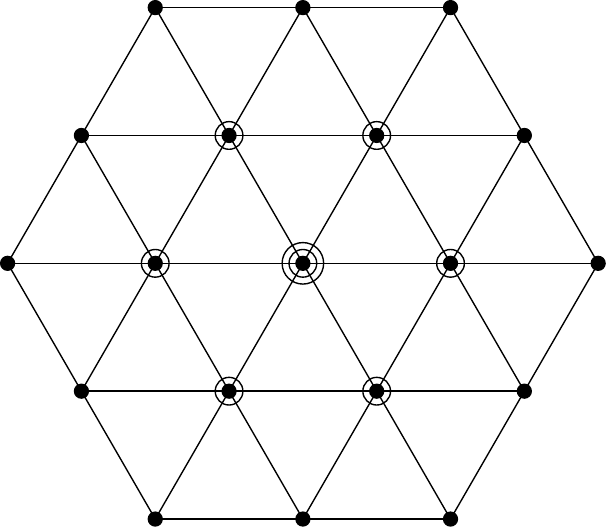}
\end{center}

\caption{(a) The weight diagram of the irreducible representation of $\mathfrak{su}(3)$ with $j=1$. This is the adjoint representation and corresponds to quantum theory. (b) The weight diagram of the irreducible representation of $\mathfrak{su}(3)$ with $j=2$. This corresponds to the simplest non-quantum state space for ${\rm P}\mathbb C^3$.}\label{weightfig}
\end{figure}

Note that quantum theory corresponds to $j=1$.
In figure~\ref{weightfig} the weight diagrams of the quantum ($\mathcal{D}_1^3$) and lowest dimensional non-quantum ($\mathcal{D}_2^3$) representations of $\mathfrak{su}(3)$ (the Lie algebra of $\mathrm{SU}(3)$) are shown. For $d=2$, $\mathcal D_j^2$ are the $\mathrm{SU}(2)$ irreducible representations with integer spin $j$, which are also irreducible representations of $\mathrm{SO}(3) \cong \mathrm{PU}(2)$.
For $d\geq 3$, these irreducible representations are also denoted with the Dynkin label $(j, \underbrace{0,\ldots,0}_{d-3} ,j)$. 

\begin{result}\label{R2}
  Let $\omega^d_j \in \mathbb R^{D_j^d}$ be the unique (up to proportionality) invariant vector $\mathcal D^d_j (U) \omega^d_j =\omega^d_j$ for all elements of the subgroup
\begin{equation}
  \label{stabilizer}
  U= \left(\begin{array}{c|c}  
    e^{i \alpha} & 0\ \ \ \cdots\ \ \ 0 
    \\ \hline
    0 & \\
    \vdots & e^{-i \alpha/(d-1)} u \\
    0 &    
  \end{array}\right),
  \quad 
  u\in {\rm SU}(d-1)\ .
\end{equation}
Each finite-dimensional representation $\Omega: \pcd \to \mathbb R^n$ and $\Gamma: \pud \to {\rm GL}(\mathbb R^n)$ satisfying~\eqref{dynam struc 2}, \eqref{dynam struc 1}, \eqref{span O} is of the form
\begin{align}
  \label{rep Gamma}
  \Gamma_U &= 
  \bigoplus_{j\in \mathcal J} 
  \mathcal D^d_j(U)\ ,
  \\ \label{rep Omega}
  \Omega_{\psi_0} &= 
  \bigoplus_{j\in \mathcal J} \omega^d_j\ ,
\end{align}
where $\mathcal J$ is any finite set of positive integers.
\end{result}

Each set $\mathcal J$ corresponds to an unrestricted theory, and clearly the dimension of the state space $n= \sum_{j\in \mathcal J} D_j^d$.
Quantum theory corresponds to $\mathcal J = \{1\}$ and $n=d^2-1$. 
Now, let us analyze a simple example: quantum theory with $d=2$.  In this case, the subgroup ${\rm SU}(d-1) = \{\unity \}$ is trivial, and the subgroup~\eqref{stabilizer} is $e^{ Z t}$ where
\begin{equation}\label{Z def}
  Z = \left( \begin{array}{cc}
    i & 0\\
    0 &-i
  \end{array}\right)\ .
\end{equation}
The action of this subgroup on the Bloch vector has two invariant states $(0,0,\pm 1)$. But, as the result says, both are the same vector up to a proportionality factor. Also note that the two state spaces generated by the two reference vectors $\omega^2_1 = (0,0,\pm 1)$ are related as in~\eqref{uni1}-\eqref{uni3}.

A consequence of Result~\ref{R2} is that all finite-dimensional group actions $\Gamma$ are polynomials of the fundamental action $U$.
Alternative~\eqref{alternative 1} is not polynomial, and hence needs an infinite-dimensional representation $V$. This shows our previous claim that an infinite number of parameters is necessary to specify a mixed state in this alternative theory.

A desirable feature of any alternative to the measurement postulates $\mathcal F_d$ is \emph{faithfulness}: for any pair pure states $\psi, \psi' \in \pcd$ there is a measurement $F\in \mathcal F_d$ which distinguish them $F(\psi) \neq F(\psi')$. When this does not happen, the two states $\psi, \psi'$ become operationally equivalent.
Faithfulness translates to the injectivity of the map $\Omega: \pcd \to V$. The following result tells us which of the representations~\eqref{rep Gamma}-\eqref{rep Omega} are faithful.

\begin{result}[Faithfulness]\label{R3}
  When $d\geq 3$ the map $\Omega$ is always injective. 
    When $d= 2$ the map $\Omega$ is injective if and only if $\mathcal J$ contains at least one odd number.
\end{result}

\noindent In this work we are specially interested in faithful state spaces. Because if two pure states in $\pcd$ are indistinguishable, then, from an operational point of view, they become the same state. 

\subsection{A simpler characterization}\label{simple}

In this section we present a different representation for all these alternative theories that makes them look closer to quantum theory. 
For this, we have to relax the ``minimality of $V$'' property (given by equation (\ref{span O})).

For any $\psi \in \pcd$ and $U\in \pud$ let
\begin{align}
  \label{qn1}
  \Omega_\psi 
  &= 
  |\psi\rangle\!\langle\psi |^{\otimes N}\ ,
  \\ \label{qn2}
  \Gamma_U \omega
  &=
  U^{\otimes N} \omega\, U^{\dagger \otimes N}\ ,
\end{align}
where $N$ is a positive integer and $\omega$ is a Hermitian matrix acting on the symmetric subspace of $(\mathbb C^d)^{\otimes N}$.
It can be seen that this representation decomposes as
\begin{equation}
  \label{qnd}
  \Gamma_U = 
  \bigoplus_{j=0}^N
  P_{N,j}^d 
  \otimes\mathcal D^d_j (U)
  \ ,
\end{equation}
where $P_{N,j}^d$ is a projector whose rank depends on $N,j,d$. The rank of projector $P_{N,j}^d$ counts the number of copies of representation $\mathcal D^d_j$. For example ${\rm rank} P_{N,0}^d =1$ and ${\rm rank} P_{N,N}^d =1$. Also, we define $\mathcal D^d_0$ to be the trivial irreducible representation.

The fact that representation~\eqref{qn1}-\eqref{qn2} contains the trivial irreducible representation, allows for effects to be linear instead of affine function. Hence, all effects can be written as
\begin{equation}
  P(E|\psi) = 
  {\rm tr} (E |\psi\rangle\!\langle\psi |^{\otimes N})
  \ ,
\end{equation}
where $E$ is a Hermitian matrix.
However, this matrix $E$ does not need to be positive semi-definite.
For example, any $N$-party entanglement witness $W$ has negative eigenvalues but gives a positive value ${\rm tr} (W |\psi\rangle\!\langle\psi |^{\otimes N}) \geq 0$ for all $\psi$.
An example of this is given in Section~\ref{3dis}.

If we want to use notation~\eqref{qn1}-\eqref{qn2} to represent the alternative theory characterized by $\mathcal J$, then we have to set $N\geq \max \mathcal J$ and constrain the allowed effects to only have support on the subspaces of~\eqref{qnd} with $j\in \mathcal J$.
The fact that for some $j\in \mathcal J$ the representations $\mathcal D^d_j$ may be repeated is irrelevant. We can restrict the effects to act on a single copy or not. In the second case the action of the effect becomes the average of its action in each copy.

\section{Phenomenology of the alternative theories}\label{sectionphenomenology}

In this section we explore the properties of the alternative theories classified in the previous section. We first consider the case $d=2$ for both irreducible and reducible representations, and later we generalize to $d \geq 3$. Before considering these families of theories we provide an example of a specific theory of $\mathrm{P}\mathbb{C}^2$ which differs significantly from quantum theory.

\subsection{Three distinguishable states in $\mathbb C^2$}\label{3dis}

\begin{result}
The $d=2$ state space $\mathcal J = \{1,2\}$ has at least three perfectly distinguishable states when all effects are allowed. 
\end{result}

\noindent (Result independently obtained in~\cite{3JB}.) The irreducible representations $\mathcal{D}_j^2$ are given by the symmetric product $\mathcal{D}_j^2 (U) = \mathrm{Sym}^{2j} U$~\cite[p.150]{Fulton91}, where $U$ is the fundamental representation on $\mathbb C^2$. According to Result~\ref{R3}, these are faithful when $j$ is odd ($\Omega$ injective). 
To prove Result~4 we first show that the unfaithful $ \mathrm P  \mathbb C^2$ theory $\mathcal J' = \{2\}$ corresponds to a quantum $\mathrm P  \mathbb C^3$ system restricted to the real numbers (i.e. $\mathrm P  \mathbb R^3$), and hence has 3 distinguishable states. By including the trivial representation, the representation $\mathcal{J}'$ can be expressed as: $\mathrm{Sym}^4 U \oplus \mathrm{Sym}^0 U = \mathrm{Sym}^2 (\mathrm{Sym}^2 U)$~\cite[p.152]{Fulton91}. Using the fact that $\mathrm{Sym}^2 U$ is the quantum representation, the states generated by applying $\mathcal{J}'$ to the reference state can be expressed as the rank-one projectors
\begin{equation}
  |\phi\rangle\! \langle\phi| = 
\mathcal{D}_1^2(U)  \omega_1^2 \  [\mathcal{D}_1^2(U) \omega_1^2]^T 
\ ,
\end{equation}
where $\omega_1^2 \in \mathbb R^3$ is a unit Bloch vector and hence $|\phi\rangle \in \mathbb R^3$ with $\langle\phi |\phi\rangle =1$. This is the state space of a 3-dimensional quantum system restricted to the real numbers $\mathrm P \mathbb R^3$, and hence, has three distinguishable states. 

Coming back to $\mathcal{J} = \{1,2\}$, this theory has transformations given by the representation $\mathrm{Sym}^4 U \oplus \mathrm{Sym}^2 U \oplus  \mathrm{Sym}^0 U$ and hence it is faithful. It has at least three distinguishable states since one can make measurements with support on the blocks $\mathrm{Sym}^4 U \oplus  \mathrm{Sym}^0 U$ to distinguish the three states. It may be the case that by making measurements with support in both blocks one could distinguish more states.
We now consider families of alternative theories with dynamics $\mathrm{SU}(2)$ and find various properties which distinguish them from the qubit.

\subsection{Irreducible $d=2$ theories}

In this section we define a class of $d=2$ theories and explore which properties they have in common with the qubit, and which properties distinguish them from quantum theory.
We denote by $\mathcal{T}_2^I$ the class of non-quantum $d=2$ theories which are faithful, unrestricted (all effects allowed) and have irreducible $\Gamma$. According to Result~\ref{R3}, each of these theories is characterized by an odd integer $j\geq 3$, such that $\mathcal J = \{j\}$, or equivalently $\Gamma = \mathcal{D}_j^2$.

\subsubsection{Number of distinguishable states}

An important property of a system is the maximal number of perfectly distinguishable states it has. This quantity determines the amount of information that can be reliably encoded in one system. For instance a classical bit has two distinguishable states, as does a qubit. Distinguishable states of a qubit are orthogonal rays in the Hilbert space, or equivalently, antipodal states on the Bloch sphere.

The following result tells us about the similarities of the theories in $\mathcal{T}_2^I$ with respect to quantum mechanics. It also applies to other theories, because it does not require the assumption of unrestricted effects.

\begin{result}\label{R4}
Any $d=2$ theory with irreducible $\Gamma$ has a maximum of two perfectly distinguishable states.
\end{result}

\noindent However all these theories have an important difference with quantum theory.

\begin{result}\label{Rnonantipodal}
All theories in $\mathcal{T}_2^I$ have pairs of non-orthogonal (in the underlying Hilbert space) rays  which are perfectly distinguishable.
\end{result}

\noindent We also prove that two rays $\psi, \phi \in {\rm P}\mathbb C^2$ are orthogonal if and only if they are represented by antipodal states $\Omega_\psi = - \Omega_\phi$. 
The existence of distinguishable non-antipodal states entails that theories in $\mathcal{T}_2^I$  have exotic properties not shared by qubits. We discuss a few of these in what follows. 

\subsubsection{Bit symmetry}

Bit symmetry, as defined in~\cite{Muller_power_2011},  is a property of theories whereby any pair of pure distinguishable states $(\omega_1 , \omega_2)$ can be mapped to any other pair pure of distinguishable states $(\omega_1' , \omega_2')$ with a reversible transformation $U$ belonging to the dynamical group, i.e. $\Gamma_U \omega_1 = \omega_1'$ and $\Gamma_U \omega_2 =  \omega_2'$. The qubit is bit symmetric since distinguishable states are orthogonal rays, and any pair of orthogonal rays can be mapped to any other pair of orthogonal rays via a unitary transformation.
\begin{result}\label{Resultbitsymmetry}
All theories in $\mathcal{T}_2^I$ violate bit symmetry.
\end{result} 
\noindent This result follows directly from Result~\ref{Rnonantipodal} and from the fact that there does not exist any reversible transformation which maps a pair of antipodal states (representations of orthogonal rays) to a pair of non-antipodal states (representations of non-orthogonal rays).  

\subsubsection{Phase invariance of measurements}

We now consider the concept of phase groups of measurements following~\cite{Garner_framework_2013, Lee_generalised_2016}. The phase group $T$ of a measurement $(E_1 , ... , E_n)$   is the maximal subgroup of the transformation group of the state space $\mathcal{S}$ which leaves all outcome probabilities unchanged: 
\begin{equation}
  E_i \circ \Gamma_U = E_i
  \ \ \forall\, U \in T \ .
\end{equation}
For example, in the case of the qubit and a measurement in the $Z$ basis, the phase group associated to this measurement is $T = \{e^{Z t }: t \in \mathbb R\}$, where~\eqref{Z def}.
We maintain the historical name ``phase group'', although in general, it need not be abelian.

A measurement which distinguishes the maximum number of pure states in a state space is often called a \textit{maximal measurement}~\cite{Garner_framework_2013}. For a qubit, these are the projective measurements (since they perfectly distinguish two states) which have a phase group $\mathrm{U}(1)$.

\begin{result}\label{phasegroupres}
All theories in  $\mathcal{T}_2^I$ have maximal measurements with trivial phase groups $T=\{\unity\}$.
\end{result}

\noindent These maximal measurements with trivial phase groups are those used to distinguish non-antipodal states. We note that the antipodal states can be distinguished by measurements with $\mathrm{U}(1)$ phase groups (as in quantum theory). We observe that contrary to \cite{Garner_framework_2013} we find maximal measurements with trivial phase groups in a non-classical theory. This is because unlike \cite{Garner_framework_2013} we do not consider all allowed transformations which map states to states, but only the transformations $\Gamma_U$. 

\subsubsection{No simultaneous encoding}

No simultaneous encoding~\cite{Masanes_existence_2013} is an information-theoretic principle which states that if a system is used to perfectly encode a bit it cannot simultaneously encode any other information (similarly for a trit and higher dimensions). More precisely, consider a communication task involving two distant parties, Alice and Bob.
Similarly as in the scenario for information causality~\cite{IC}, suppose that Alice is given two bits $a,a' \in \{0,1\}$, and Bob is asked to guess only one of them. He will base his guess on information sent to him by Alice, encoded in one $\mathcal{T}_2^I$ system. Alice prepares the system with no knowledge of which of the two bits, $a$ or $a'$, Bob will try to guess.
No simultaneous encoding imposes that, in a coding/decoding strategy in which Bob can guess $a$ with probability one, he knows nothing about $a'$. That is, if $b,b'$ are Bob's guesses for $a,a'$ then
\[
	P(b|a,a')= \delta^a_b  \Rightarrow 
	P(b'|a,a'=0)= P(b'|a,a'=1)
\]
where $\delta^a_b$ is the Kronecker tensor. 

As an example consider a qubit. Alice decides to perfectly encode bit $a$, which she can only do by encoding $a = 0$ and $a=1$  in two perfectly distinguishable states. Without loss of generality she can choose to encode $a = 0$  in  $\ket{0}$ and $a = 1$ in $\ket{1}$, with $\langle 0 |1\rangle =0$. She now also needs to encode $a'$ whilst keeping $a$ perfectly encoded. Since, $\ket 0$ is the only state which is perfectly distinguishable from $\ket 1$, we have that both $a = 0 , a' = 0$ and $a = 0 , a'=1$ combinations must be assigned to $\ket{0}$. Similarly $a = 1 , a' = 0$ and $a = 1 , a'=1$ combinations must be assigned to $\ket{1}$. In this case we see that whilst Bob can perfectly guess the value of $a$ if he chooses to, if he chooses to guess the value of $a'$ he cannot do so.
Hence this property is met by qubits, however

\begin{result}\label{Rnosim}
All theories in $\mathcal{T}_2^I$ violate no simultaneous encoding.
\end{result} 

\noindent We see that the above three properties: bit symmetry, existence of phase groups for maximal measurements and no-simultaneous encoding single out the qubit amongst all $\mathcal{T}_2^I$ theories.

\subsubsection{Restriction of effects}

The study of theories in $\mathcal{T}_2^I$ has shown that they differ from quantum theory in many ways. We now consider a new family of theories $\tilde{\mathcal{T}}_2^I$, which is constructed by restricting the effects of the theories in $\mathcal{T}_2^I$.
These theories turn out to be closer to quantum theory in that they obey all the above properties. This approach is similar to the self-dualization procedure outlined in~\cite{Janotta_gen_2013} (which also recovers bit symmetry).

We call a theory \textit{pure-state dual} if the only allowed effects are ``proportional'' to pure states. That is, for every allowed effect, there is a pure state $\psi$ and a pair of normalization constants $\alpha, \beta$ such that 
\begin{equation}
  E (\omega) = 
  \alpha (\Omega_{\psi}^T \cdot \omega) 
  +\beta\ .
\end{equation}
All theories in $\tilde{\mathcal{T}}_2^I$ have a maximum of two distinguishable states, and all pairs of distinguishable states are antipodal.

\begin{result}\label{PSDres}
All theories in $\tilde{\mathcal{T}}_2^I$ are bit symmetric, have phase groups for all maximal measurements and obey no-simultaneous encoding.
\end{result}

\subsection{Reducible $d=2$ theories}

We consider the set of all unrestricted faithful non-quantum state spaces generated by reducible representations of $\mathrm{SU}(2)$. These representations are given by equation~\eqref{rep Gamma} with $d=2$ and  $|\mathcal{J}|>1$ containing at least one odd number. We denote the set of all these theories $\mathcal{T}_2^R$.

For theories in $\mathcal{T}_2^R$ the number of distinguishable states can be more than two (as shown for $\mathcal{J} = \{1,2\}$ in Section~\ref{3dis}). It is in general a difficult task to find the number of distinguishable states. However we find:

\begin{result}\label{Bitsymred}
All theories in $\mathcal{T}_2^R$ violate bit symmetry.
\end{result}
 
That is, there are pairs of distinguishable pure states $(\psi_1 , \psi_2)$ which cannot be mapped to another pair of distinguishable states $(\psi_1' , \psi_2')$ with a reversible transformation $U$ belonging to the dynamical group.
This implies that 

\hspace{1cm}

\begin{framed}
\noindent \emph{Bit symmetry singles out quantum theory amongst all $d=2$ unrestricted faithful state spaces (both irreducible and reducible).}
\end{framed}

\begin{figure}
\includegraphics[ width = \columnwidth]{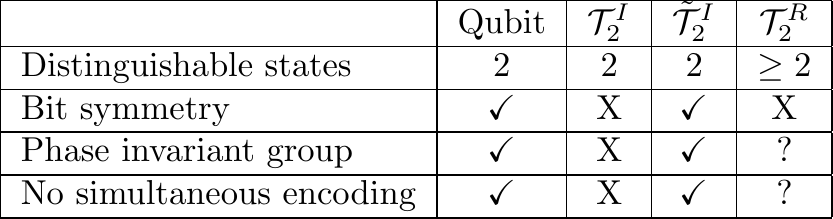}
	\caption{Summary of results for $d=2$ theories. 
	}\label{sumtable}
\end{figure}

\subsection{Arbitrary $d$}

Having studied in detail different families of $d=2$ theories, we now consider properties of arbitrary-$d$ unrestricted theories. We also show that the property of bit symmetry singles out quantum theory in all dimensions.  This is proven by showing that all faithful $\pcd$ state spaces have embedded within them faithful (reducible) $\mathrm P \mathbb C^{d-1}$ state spaces. We then show that if a state space $\mathrm P \mathbb C^{d-1}$ violates bit symmetry, then any state space $\pcd$ it is embedded in also does. Using a proof by induction we find:
\begin{result}\label{resultgend}
All unrestricted non-quantum theories with pure states $\pcd$ and transformations $\mathcal{D}_j^d$ are not bit symmetric.
\end{result}
\noindent This result shows that the Born rule can be singled out amongst all possible probability assignments from the following set of assumptions.

\hspace{1cm}

\begin{framed}
\noindent \emph{The Born rule is the unique probability assignment satisfying: (i) no restriction on the allowed effects, and (ii) bit symmetry.}
\end{framed}

\hspace{1cm}

\noindent Although many works in GPT's only consider unrestricted theories  no-restriction remains an assumption of convenience with a less direct physical or operational meaning. Bit symmetry however is a property which has computational and physical significance as discussed in~\cite{Muller_power_2011}. It is generally linked to the possibility of reversible computation, since bit symmetric theories allow any logical bit (pair of distinguishable states) of the theory to be reversibly transformed into any other logical bit.

\section{Conclusion}\label{sectionconclusion}

In this work we have considered theories with the same structure and dynamics for pure states as quantum theory but different sets of OPF's (not necessarily following the Born rule). These alternative theories were shown to be in correspondence with certain class of representations of the unitary group. Moreover these theories (assuming no restriction on the effects) were shown to differ from quantum theory in that they are not bit symmetric. In the case of $d=2$, the number of distinguishable states was bounded for irreducible representations, and further properties which single out quantum theory were found. 

In order to fully develop these alternative theories it is necessary to determine how the state spaces studied compose. This would then allow us to study important properties related to multipartite systems.

Gleason's theorem~\cite{Gleason_measures_1975} can be understood as a derivation of the Born rule.	Gleason assumes that measurements are  associated with orthogonal bases and shows that states must be given by density matrices and measurement outcome probabilities by the Born rule. This is a different approach to the one presented in this paper. 
In this work we show that, assuming that pure states are rays on a Hilbert space and time evolution is unitary, the Born rule can be derived (or singled out) from the requirement of bit symmetry. We allow for measurements to take any form, not just being associated with orthogonal bases in the corresponding Hilbert space. Unlike Gleason's theorem this applies in the case $d=2$ and may arguably provide a more physical justification for the Born rule. There exists a derivation of Gleason's theorem for POVM's  \cite{Caves_gleason_2004} which is closer in spirit to our approach than the original. However we emphasise that in our work we begin from the structure of pure states and dynamics to derive that of measurements. In Gleason's theorem (and its extension) it is the structure of measurements which are assumed and that of states derived from it.

Future work will be focused on multipartite systems and post-measurement state update rules for the alternative theories analysed in this work. This will allow us to consider more general information processing tasks for these novel probabilistic theories.

\section{Acknowledgements}

We are grateful to Howard Barnum, Jonathan Barrett, Markus Mueller, Jonathan Oppenheim, Matthew Pusey, Jonathan Richens and Tony Short for valuable discussions. We acknowledge the helpful referee comments which allowed us to correct the proofs of Lemma~2 and Result~7 as well as clarify some technical points about the continuity of the OPF's. TG is supported by the EPSRC Centre for Doctoral Training in Delivering Quantum Technologies. LM is supported by the EPSRC.

\bibliographystyle{unsrtnatmod}
\bibliography{refs}

\appendix

\section{Proofs}

\subsection{The convex representation (proof of Result~\ref{R1})}\label{convexappendix}

Let us fix the value of $d$ and the set of OPFs $\F_d$.
Also, we assume that all OPFs of the form $F\circ U$ are contained in $\F_d$, for all $U\in \pud$ and $F \in \F_d$. And recall that, if this is not the case, we can always re-define $\F_d$ to include them.
Before defining the representation $(\Omega, \Gamma, \Lambda)$ of Result~\ref{R1}, it is convenient to define a temporary one $(\tilde\Omega, \tilde\Gamma, \tilde\Lambda)$ in which states $\tilde\Omega _\psi$ are represented by a list of outcome probabilities.
For this, we take a minimal subset $\{F^{(1)}, F^{(2)}, F^{(3)}, \ldots\} \subseteq \F_d$ that affinely generates $\F_d$. That is, for any $F \in \F_d$ there are constants $c^{(0)}, c^{(1)}, c^{(2)}, \ldots \in \mathbb R$ such that $F = c^{(0)} + \sum_r c^{(r)} F^{(r)}$. 
This construction requires the minimal generating subset $\{F^{(1)}, F^{(2)}, \ldots\}$ to be countable, but it can still be infinite.
This does not pose any loss of generality, because, from an operational point of view, the characterization of arbitrary states would be impossible.

If we represent the pure state $\psi$ by the list of probabilities
\begin{equation}
  \tilde \Omega_{\psi} 
  = 
  \left(\begin{array}{c}
  P(F^{(1)} |\psi) \\ 
  P(F^{(2)} |\psi) \\ 
  P(F^{(3)} |\psi) \\
  \vdots 
\end{array} \right) 
= 
  \left(\begin{array}{c}
  F^{(1)} (\psi) \\ 
  F^{(2)} (\psi) \\ 
  F^{(3)} (\psi) \\
  \vdots 
\end{array} \right) 
  \in V \ ,
\end{equation} 
then we can use the rules of probability to extend this definition to mixed states. For example, the probability of outcome $F$ when the system is prepared in the ensemble $(p_i, \psi_i)$ is
\begin{equation}
  P(F|(p_i, \psi_i)_i) 
  =
  \sum_i p_i P(F|\psi_i) 
  =
  \sum_i p_i F (\psi_i) \ .
\end{equation}
Therefore, the probabilistic representation of ensemble $(p_i, \psi_i)$ is $\sum_i p_i \tilde\Omega _{\psi_i}$. And the set of all states, pure and mixed, is
\begin{equation}
  \label{span tilde}
  \tilde {\cal S} = 
  {\rm conv}\, \tilde \Omega _{\pcd}
  \ .
\end{equation}

In the representation $\tilde \Omega$, any probability $F (\psi)$ is either a component of the vector $\tilde \Omega_\psi$ or an affine function of its components. Hence, for any $F$, there is an affine function $\tilde \Lambda_F : V \to \mathbb R$ such that
\begin{equation}
  \label{cond tilde 1}
  \tilde\Lambda _F (\tilde\Omega _\psi)
  = F(\psi)
  \quad \forall\, \psi\ .
\end{equation}

Any $U$ can be seen as a relabeling of the pure states $\pcd$, hence, $\{(F^{(1)} \circ U), (F^{(2)} \circ U), \ldots\} \subseteq \F_d$ is also a minimal generating subset. 
Therefore, there is an affine map $\tilde \Gamma_U: V \to V$ such that 
\begin{equation}
  \label{cond tilde 2}
  \tilde\Omega_{U\psi}
  =
  \tilde\Gamma_U (\tilde\Omega_\psi) 
  \quad \forall\, U, \psi\ .
\end{equation}

Next, we define a new representation $\Omega, \Gamma, \Lambda$ in which $\Gamma_U:V\to V$ is linear. This new representation is that of Result~\ref{R1}.
For this, we need to define the maximally mixed state
\begin{equation}
  \tilde \omega_{\rm mm} 
  = 
  \int_{\pud} 
  \hspace{-6mm} dU\,
  \tilde \Gamma_U (\tilde \Omega_{\psi})
  \ ,
\end{equation}
where $dU$ is the Haar measure and $\psi$ is any pure state. 
We also note that the maximally mixed state is invariant under any unitary: $\tilde \Gamma_U (\tilde \omega_{\rm mm}) = \tilde \omega_{\rm mm}$ for any $U$.
Now, we define the new representation as
\begin{align}
  \label{omega to tilde}
  \Omega_\psi &= \tilde \Omega_\psi 
  -\tilde \omega_{\rm mm} 
  \ , \\
  \Gamma_U (\omega) &=
  \tilde \Gamma_U (\tilde \omega)   -\tilde \omega_{\rm mm} 
  \ , \\
  \Lambda_F (\omega) &=
  \tilde \Lambda_F (\tilde \omega ) 
  \ ,
\end{align}
which extends to general mixed states as $\omega = \tilde \omega - \tilde \omega_{\rm mm}$.
Using~\eqref{cond tilde 1} and~  \eqref{cond tilde 2} we obtain~\eqref{dynam struc 2}, \eqref{dynam struc 1}.
In this representation, the maximally mixed state is the zero vector $\omega_{\rm mm} =0 \in V$. Now, recalling that $\omega_{\rm mm}$ is invariant under unitaries, we have that $\Gamma_U(0) =0$, which together with the affinity of $\Gamma_U$ implies that $\Gamma_U:V\to V$ is linear, for all $U$.

To continue with the proof of Result~\ref{R1}, we note that the affinity of $\tilde \Lambda_F$ implies that of $\Lambda_F$; and that~\eqref{span tilde} implies~\eqref{span O}.

Now, it only remains to prove uniqueness.
Let us suppose that there are other maps $\Omega', \Gamma', \Lambda'$ with the properties stated in Result~\ref{R1}. 
These properties imply the existence of an affine function $\tilde L$ such that \begin{equation}
  \tilde \Omega_{\psi} 
  = 
  \left(\begin{array}{c}
  F^{(1)} (\psi) \\ 
  F^{(2)} (\psi) \\ 
  \vdots 
\end{array} \right) 
  = 
  \left(\begin{array}{c}
  \Lambda'_{F^{(1)}}(\Omega'_{\psi}) \\ 
  \Lambda'_{F^{(2)}}(\Omega'_{\psi}) \\ 
  \vdots 
\end{array} \right) 
  = \tilde L(\Omega'_\psi)\ .
\end{equation} 
Which in turn implies the existence of another affine function $L$ such that $\Omega_\psi = L(\Omega'_\psi)$.
According to the properties stated in Result~\ref{R1}, both $\Gamma_U$ and $\Gamma'_U$ are linear; which implies that the maximally mixed state in both representations is the zero vector $\omega_{\rm mm} = \omega'_{\rm mm} =0 \in V$. 
The affine map $L$ also changes the representation of the maximally mixed state $\omega_{\rm mm} = L(\omega'_{\rm mm})$. But $0=L(0)$ is only possible if $L$ is not just affine, but linear.

All the results of this section are valid when $V$ is finite- and infinite-dimensional. In the rest of the appendices, only the finite-dimensional case is considered.

\subsubsection{Continuity of homomorphisms $\Gamma$}
\label{continuity Gamma}

The number of parameters that define a mixed state in $\mathcal S$ is equal to the dimension of $V = {\rm conv}\, \mathcal S$. Hence, the requirement that general mixed states in $\mathcal S$ can be estimated with finite means implies that $V$ is finite-dimensional.
Now, we recall that in the probabilistic representation $\tilde\Omega$, the entries of states $\tilde \omega \in \tilde {\mathcal S}$ are bounded. Hence, due to~\eqref{omega to tilde}, the same is true in $\mathcal S$.
This implies that the absolute value of all matrix elements of the group $\Gamma_{\pud}$ are also bounded. 

Now, we argue that, from a physical point of view, the group of transformations $\Gamma_{\pud}$ must be topologically closed. 
This follows from the fact that any mathematical transformation that can be approximated arbitrarily well by physical transformations should be a physical transformation too.
In summary, the fact that the set $\Gamma_{\pud} \subseteq \mathbb R^n$  is bounded and closed implies that it is compact.
And this is the premise of the following theorem.

It is proven in Theorem~5.64 of~\cite[p.167]{Hofman_structure_2013} that any group homomorphism $\Gamma: \pud \to \mathcal G$ in which $\mathcal G$ is compact must be continuous.

\subsection{Preservation of dynamical structure (Result~\ref{R2})}

The stabilizer subgroup $\mathcal G_{\psi} \leq \pud$ of a pure state $\psi\in \pcd$ is the set of unitaries $U$ leaving the state invariant $U\psi = \psi$. 
If the state $\psi'$ is related to $\psi$ via $\psi' = U\psi$, then their corresponding stabilizer subgroups are related via $\mathcal G_{\psi'} =  U\mathcal G_{\psi} U^{\dagger}$.
Therefore, all stabilizers are isomorphic.

Equation~\eqref{dynam struc 1} implies that $\Gamma$ is a representation of $\mathrm{SU} (d)$. Following the premises of Result~\ref{R2}, this representation is finite-dimensional. Hence, we can decompose $\Gamma$ into real irreducible representations
\begin{equation}
  \label{reduc}
  \Gamma = \bigoplus_r \Gamma^r\ .
\end{equation}
Using the same partition into real linear subspaces, we also decompose the map
\begin{equation}
  \Omega = \bigoplus_r \Omega^r\ .
\end{equation}
Equation~\eqref{dynam struc 2} independently holds for each summand $\Gamma^r_U \Omega^r_{\psi} = \Omega^r_{U\psi}$.
In particular $\Gamma^r_U \Omega^r_{\psi} = \Omega^r_{\psi}$ for all $U\in \mathcal G_{\psi}$, which implies that each $\Gamma^r$ has a $\mathcal G_{\psi}$-invariant subspace.

Now, let us concentrate on the $d\geq 3$ case.
To figure out the structure of the stabilizer subgroup, we consider the state $\psi_0 = (1,0,\ldots, 0)$, and note that the group $\mathcal G_{\psi_0}$ is the set of unitaries of the form~\eqref{stabilizer}. 
Hence, when $d\geq 3$, all stabilizers are isomorphic to $\mathrm U (1) \times \mathrm{SU} (d-1)$.
%
%
\begin{lemma}\label{branchingrule}
The finite-dimensional irreducible representations of $\mathrm{SU} (d)$ that have $\mathrm U (1) \times \mathrm{SU} (d-1)$-invariant vectors are the $\mathcal D^d_j$ introduced above. Additionally, the vector is always unique (up to a constant).
\end{lemma}

\noindent This lemma, proven in  appendix \ref{Lemma1proof}, tells us that most of the irreducible representations of $\mathrm{SU} (d)$ do not have a vector which is invariant under $\mathrm U (1) \times \mathrm{SU} (d-1)$, and hence violate~\eqref{dynam struc 2}.
As mentioned above, the $\mathrm{SU} (d)$ representations $\mathcal D^d_j$ are also $\pud$ representations, which is what is required (see Result~\ref{R1}). A technical point (fully addressed in appendix \ref{ReCoirrep}) arises. The representations we are looking for act on real vector spaces, however the representation theory tools used to find the $\mathcal D^d_j$ deal with representations acting on complex vector spaces. A representation which is irreducible on a real space is not necessarily irreducible on a complex space. We have found all complex irreducible representations (which are also real irreducible) with the correct properties, however it may be the case that there are real irreducible representations which are complex reducible which were not found with the methods used. It is shown in appendix  \ref{ReCoirrep} that the $\mathcal D^d_j$  are the only real irreducible representations with the required properties. 

A very convenient fact is that there is a single one-dimensional invariant subspace. Hence, the maps $\Omega^r$ are completely determined up to a factor.
However, this factor does not play any role, since it can be modified via an equivalence transformation $L$ of the type~\eqref{uni1}-\eqref{uni3}.
If there were multiple trivial invariant subspace, one could construct different alternatives to the measurement postulates having the same representation $\Gamma$.

In the case $d=2$, the stabilizer subgroup $\mathcal G_\psi$ is isomorphic to $\mathrm U (1)$. This subgroup is generated by a single Lie algebra element (e.g. $Z$). The irreducible representations $\mathcal D$ in which, the subgroup $\mathcal G_\psi$ has an invariant vector, are those in which $\mathcal D (Z)$ has at least one zero eigenvalue. These are the integer spin representations. Also, the multiplicity of this eigenvalue is always one, implying that, as in the $d\geq 3$ case, the map $\Omega$ is completely determined up to equivalence transformations of the type~\eqref{uni1}-\eqref{uni3}.

\subsection{Faithfulness (Result~\ref{R3})}

Let us prove that when $d\geq 3$ the map $\Omega$ specified in Result~\ref{R2} is injective.
We start by assuming the opposite: there are two different pure states $\psi\neq \psi'$ which are mapped to the same vector $\Omega_\psi = \Omega_{\psi'}$. This vector must be invariant under the action of the two stabilizer subgroups: $\Gamma_U \Omega_\psi = \Omega_\psi$ for all $U\in \mathcal G_\psi$ and all $U\in \mathcal G_{\psi'}$.
Now, note that if $\Gamma_U \Omega_\psi = \Gamma_{U'} \Omega_\psi = \Omega_\psi$ then also $\Gamma_{UU'} \Omega_\psi = \Omega_\psi$. Hence, the stabilizer group of the vector $\Omega_\psi$ contains the group $\{UU'| \forall\, U\in \mathcal G_{\psi}, U'\in \mathcal G_{\psi'}\}$. 
\begin{lemma}\label{subgrouplemma}
When $d \geq 3$ the group $\mathrm{SU}(d)$ can be generated by two stabilizer subgroups $\mathrm{SU}(d-1) \times \mathrm{U}(1)$ of two distinct rays on $\mathrm{P}\mathbb{C}^d$.
\end{lemma}

\noindent
This lemma, proven in Appendix \ref{Lemma2proof}, shows that the group generated by any two stabilizer subgroups is the full group $\mathrm{SU} (d)$.
Therefore, the vector $\Omega_\psi$ is invariant under any transformation $\Gamma_U$. This implies that all states $\psi'$ are mapped to the same vector $\Omega_{\psi'} = \Omega_\psi$ and have exactly the same outcome probabilities.
Equivalently, all functions $F\in \F_d$ are constant.

Now, let us analyze the $d=2$ case.
In what follows we prove that, the map $\Omega^r$ associated to an irreducible representation $\Gamma^r = \mathcal D^2_j$ is injective when $j$ is odd. 
Hence, the global map $\Omega$ is injective if it contains at least one summand $\Omega^r$ that is injective. 
An irreducible representation of $\mathrm{SU}(2)$ can be expressed as a symmetric power of the fundamental representation \cite[p.150]{Fulton91}:
\begin{equation}
\mathcal D^2_j = \mathrm{Sym}^{(2j)} \mathcal D^2_{\frac{1}{2}} \ ,
\end{equation}
where $ D^2_{\frac{1}{2}}$ is the fundamental representation of $\mathrm{SU}(2)$ acting on $\mathbb{C}^2$, with basis $\{\psi_0,\psi_1\}$. The action of the Lie algebra element $Z$ on this basis is:
\begin{align}
Z \psi_0 &= i \psi_0 \ ,\\
Z \psi_1 &= - i \psi_1 \ .
\end{align}
Given $\mathcal D^2_j$, we take as reference state $\Omega^j_{\psi_0}$,  the 0 eigenstate of $\mathcal D^2_j(Z)$ (since $\psi_0$ is invariant under all transformations generated by $Z$).
\begin{equation}
\mathrm{e}^{ \mathcal D^2_j(Z) t} \Omega^j_{\psi_0} = \Omega^j_{\psi_0} \ .
\end{equation}
The 0 eigenstate is given by  $\Omega^j_{\psi_0} = \psi_0^{\otimes j} \otimes  \psi_1^{\otimes j} $ (where the product is the symmetric product)~\cite[p.150]{Fulton91}. All states can be obtained by applying a unitary to the reference state: $\Omega^j_{\psi} = \mathcal D^2_j (U) \Omega^j_{\psi_0}$. 
We call $U_Z$ the set of transformations generated by $Z$. All states $\psi \neq \psi_0$ are of the form $\psi = U \psi_0, \ U \notin U_Z$.

We show that for $j$ odd there are no states $ U \psi_0 ,\ U \notin U_Z $ such that $\mathcal D^2_j(U) \Omega^j_{\psi_0} = \Omega^j_{\psi_0}$ hence $\psi_0$ and $\psi$ are mapped to distinct states. For $j$ even we show that there is a $U \notin U_Z$ such that $\Omega^j_{\psi} = D^2_j(U) \Omega^j_{\psi_0} = \Omega^j_{\psi_0}$ and hence the representation is not faithful.
A generic $U \in \mathrm{SU}(2)$ acting on $\mathbb{C}^2$ has the following action:
\begin{align}
\psi_0 &\rightarrow \alpha \psi_0 + \beta \psi_1 \nonumber \ , \\
\psi_1 &\rightarrow \alpha^* \psi_1 - \beta^*\psi_0 \ ,
\end{align}
where $| \alpha|^2 + |\beta|^2 = 1$.
The action $\mathcal D^2_j(U)$ on $\Omega^j_{\psi_0}$ is the same as that of $U^{\otimes 2j}$ since  $\Omega^j_{\psi_0}$ belongs to the symmetric subspace.
\begin{align}
\psi_0^{\otimes j} &\rightarrow (\alpha \psi_0 + \beta \psi_1)^{\otimes j} \nonumber \ , \\
\psi_1^{\otimes j} &\rightarrow (\alpha^* \psi_1 - \beta^*\psi_0)^{\otimes j} \ .
\end{align}
We now determine which unitaries $U$ preserve the state:
\begin{equation}
\psi_0^{\otimes j} \psi_1^{\otimes j} = (\alpha \psi_0 + \beta \psi_1)^{\otimes j} (\alpha^* \psi_1 - \beta^*\psi_0)^{\otimes j} \ .
\end{equation}
This only holds when either $\alpha$ or $\beta$ is 0. When $\beta = 0$ this corresponds to a $U \in U_Z$. When $\alpha=0$ we have:
\begin{equation}
\psi_0^{\otimes j} \psi_1^{\otimes j} = (-1)^j  \psi_0^{\otimes j} \psi_1^{\otimes j} \ ,
\end{equation}
since by unitarity requirement $|\beta| = 1$.
For even $j$ there is a unitary $U \notin U_Z$, such that $\mathcal D^2_j(U) \Omega^j_{\psi_0} = \Omega^j_{\psi_0}$ hence the map $\Omega$ is not injective. It maps orthogonal rays to the same state. For odd $j$ any unitary $U \notin U_Z$ maps $\Omega^j_{\psi_0}$ to a different state and so the map $\Omega$ is injective. Moreover we see that $U \psi_0 = \psi_1$ and hence orthogonal rays $\psi_0$ and $ \psi_1$ are mapped to antipodal states $\Omega_{\psi_0}$ and $\Omega_{\psi_1} = -\Omega_{\psi_0}$ for odd $j$ and to the same state $\Omega_{\psi_0} = \Omega_{\psi_1}$ for even $j$.

\subsection{Distinguishable states in $\mathcal{T}_2^I$ (Result~\ref{R4})}

The proof is by contradiction: we assume the existence of 3 distinguishable states, and show that any distinguishing measurement has an outcome with a negative probability for certain states. Hence the maximum number of perfectly distinguishable states is two. Since effects are affine functions we can write an effect $E$ as a pair $(\mathbf{e}, c)$.
\begin{equation}
E(\omega) = \mathbf{e} \cdot \omega + c   \ .
\end{equation}
A measurement is a set of effects $\lbrace E_i \rbrace$ such that $\sum E_i (\omega) = 1$ for all states $\omega$; this entails $\sum \mathbf{e}_i = \mathbf{0}$ and $\sum c_i = 1 $. If three states $\omega_1$, $\omega_2$ and $\omega_3$ are distinguishable then there exists a measurement which distinguishes them of the following form:  $\lbrace E_1 = ( c_1 , \be_1 ) ,E_2 = ( c_2 , \be_2 ),E_3= ( c_3 , \be_3 ) \rbrace$, with $E_i( \omega_j) = \delta_{ij}$. This entails 
\begin{equation}
\be_i \cdot \omega_i = 1 - c_i \ .
\end{equation}
From Result~\ref{R3} there exists an antipodal state $- \omega_i$ for every $\omega_i$ in theories in $\mathcal{T}_2^I$. We compute the outcome probabilities for these antipodal states:
\begin{equation}
E_i(-\omega_i) = \be_i \cdot (- \omega_i) + c_i = -1 + 2 c_i
\end{equation}
The sum of these three measurement outcomes probabilities is:
\begin{equation}
\sum_i E_i(-\omega_i) = -3 + 2 (c_1 +c_2 +c_3) = -1
\end{equation}
Therefore at least one of the outcome probabilities is negative and therefore not legitimate. By contradiction this proves there is no measurement which perfectly distinguishes three states in these theories.

\subsection{Distinguishable non-antipodal states in $\mathcal{T}_2^I$ (Result~\ref{Rnonantipodal})}
\label{nonantstates}

In this section we show that for each non-quantum theory in $\mathcal{T}_2^I$  there exists a measurement of the form $M = \{(\mathbf{e},c),(-\mathbf{e},1-c)\}$ with $\mathbf{e} \propto \omega_0^\dagger \cdot \mathcal{D}_j^2(X)^\dagger$ which perfectly distinguishes pairs of non antipodal states lying on the $\mathcal{D}(U_X(t)) \omega_0$ orbit. 

Theories in $\mathcal{T}_2^I$ are generated by representations $\mathcal{D}_j^2$ with $j$ odd. The dimension of such representations is $n=2j+1$. We write $U_K(t) = e^{K t/2}$ for an arbitrary element of the Lie Algebra $K$. We will use the following elements of $\mathfrak{su}(2)$:
\begin{equation}
Z = \left(
\begin{array}{cc}
i & 0 \\
0 & -i
\end{array} \right) \ ,
\end{equation}
\begin{equation}
X = \left(
\begin{array}{cc}
0 & i \\
i & 0
\end{array} \right) \ ,
\end{equation}
\begin{equation}
H = \frac{1}{\sqrt{2}}\left(
\begin{array}{cc}
i & i \\
i & -i
\end{array} \right) \ ,
\end{equation}
We note that $\mathcal{D}_j^2(e^{K t/2}) = e^{ \mathcal{D}_j^2(K) t}$. 
The reference state is chosen to be invariant under all $U_Z(t)$:
\begin{equation}
\mathcal{D}_j^2(U_Z(t))\omega_j^2 = \omega_j^2 \ ,
\end{equation} 
which implies $\mathcal{D}_j^2(Z)\omega_j^2 = 0$. The $0$-weight subspace is one dimensional, the other possible (normalised) state is $-\omega_j^2$. This will be the antipodal state. 
The manifold of pure states can be generated in the following manner:
\begin{equation}
\omega(s,t) = \mathcal{D}_j^2(U_Z(s)) \mathcal{D}_j^2(U_X(t))\omega_j^2 \ .
\end{equation}
This is just a 2-sphere (embedded in a space of dimension $n$) parametrised by polar angle $s$ and azimuthal angle $t$. The generators of the irreducible (unitary) representations of $\mathrm{SU}(2)$ can be written as \cite[p.387]{Bengtsson_Geometry_2006}:
\begin{widetext}
\begin{equation}
\begin{split}
 \mathcal{D}_j^2(X)  = 
 i \left(
	\begin{array}{ccccccc}
		 0 & \frac{\sqrt{2j}}{2} & 0 & 0 & 0 & 0 & 0 \\
		\frac{\sqrt{2j}}{2} & 0 & \frac{\sqrt{2(2j-1)}}{2} & 0 & 0 & 0 & 0 \\
		 0 & \frac{\sqrt{2(2j-1)}}{2} & 0 & \ddots & 0 & 0 & 0 \\
		 0 & 0 & \ddots & 0 & \ddots & 0 & 0 \\
		 0 & 0 & 0 & \ddots & 0 & \frac{\sqrt{2(2j-1)}}{2} & 0 \\
		 0 & 0 & 0 & 0 & \frac{\sqrt{2(2j-1)}}{2} & 0 & \frac{\sqrt{2j}}{2} \\
		 0 & 0 & 0 & 0 & 0 & \frac{\sqrt{2j}}{2} & 0 \\
	\end{array}
		\right)
\end{split} 
\end{equation}

\begin{equation}
	\mathcal{D}_j^2(Z) = i \left(
	\begin{array}{ccccccc}
	 j & 0 & 0 & 0 & 0 & 0 & 0 \\
	 0 & j-1 & 0 & 0 & 0 & 0 & 0 \\
	 0 & 0 &  \ddots & 0 & 0 & 0 & 0 \\
	 0 & 0 & 0 & 0 & 0 & 0 & 0 \\
	 0 & 0 & 0 & 0 & \ddots & 0 & 0 \\
	 0 & 0 & 0 & 0 & 0 & -j+1 & 0 \\
	 0 & 0 & 0 & 0 & 0 & 0 & -j \\
	\end{array} 
	\right) \ .
\end{equation}
\end{widetext}

We define $k=j+1$ where $\mathcal{D}(K)_{(k,k)}$ is the central element of the matrices $\mathcal{D}(K)$ for some Lie algebra element $K$. It is clear that in this basis the reference state (0 eigenstate of $\mathcal{D}_j^2(Z)$) is:
\begin{equation}
	(\omega_j^2)_{i} = \delta_{ik} \ .
\end{equation}
In the following we drop the explicit reference to the representation, and use $\mathcal{D}(U)$ for $\mathcal{D}_j^2(U)$. We now use the notation $\omega_{k+i}^K$ for the $i$ eigenstate of the   representation matrix  $\mathcal{D}(K)$ of the Lie Algebra element $K$. Here $i$ runs from $-j$ to $j$. In the case of the diagonal operator $\mathcal{D}(Z)$ we observe that an eigenvector with eigenvalue $i$ has a single entry at position $k+i$. 
The zero eigenstate  $(\omega_k^Z)_m = \delta_{km}$ which is just a vector with a single entry at position $k$ (the central entry). The 0 eigenstate of the operator $\mathcal D(X)$ for example is $\omega_k^X$.  Hence we use $\omega_k^Z$  for the reference state $\omega_j^2$ (since it is the 0 eigenstate of the Z operator). The 0 eigenstate of $\mathcal{D}(X)$  corresponds to the state $\omega_k^X$ which is equal to $\mathcal{D}(U_Y(-\frac{\pi}{2}))  \omega_k^Z$ (and is invariant under $\mathcal{D}(U_X(t))$). These 0 eigenstates of Lie Algebra matrices $\mathcal D(K)$ correspond to states since they have the correct invariance properties, that is, invariant under $\mathrm{e}^{i \mathcal{D}(K) t}$.

As shown in the proof of Result~\ref{R3} these state spaces have antipodal states for all states. Consider an effect which has as maximum $\omega_m$ will have as minimum $-\omega_m$. This effect may be noisy (i.e. not have values $0$ and $1$ for the minimum and maximum) but will be proportional to an effect which is tight (gives values 0 and 1 on the minimum and maximum). Hence any tight effect which has as maximum $\omega_m$ will distinguish it from $-\omega_m$. In the following we show that there exist effects which have 2 global maxima and minima when applied to the manifold of pure states. These effects are proportional to tight effects which distinguish $\omega_0$ and $-\omega_0$ and also distinguish $\omega_0$ and $\omega_1$ (for $\omega_1 \neq -\omega_0$).
	
Since effects $(\mathbf{e},c)$ are of the form $\mathbf{e} \cdot \omega(s,t) + c$, we need only consider $ \mathbf{e} \cdot \omega(s,t)$ to establish the number of global extrema and their locations. 
We now prove the effect $(\mathbf{e},c)$ with $\mathbf{e}=  (\omega_k^{Z})^{\dagger} \mathcal{D}(X)^\dagger$ (up to normalisation) perfectly distinguishes non-antipodal states in all representations. Here we use the $\dagger$ since the specific representation of the states have complex entries (even though the vector space spanned by the states is a real vector space). 
We first show that $\mathbf{e}$ gives real values when applied to the state space (whose states span a real vector space). There is a basis for the representation where all entries are real which is related to this representation by $A \mathcal{D}(U) A^{-1}$. It acts on states as $A \omega$ and effects as $\mathbf{e} A^{-1}$. Hence $  (\omega_j^{Z})^{\dagger} A^{-1} \cdot A \mathcal{D}(U) A^{-1} A \omega$ is real for all $\omega$ (since it is an inner product of vectors with real entries). This is equal to $ \mathbf{e} \cdot \omega$ for all omega. Hence $\mathbf{e}$ applied to the state space is real valued.

We study the maxima and minima of $f(s,t) = \mathbf{e} \cdot \omega(s,t)$ which corresponds to applying $\mathbf{e}$ to the entire manifold of pure states:
\begin{equation}
	f(s,t) = \mathbf{e} \cdot \mathcal{D}(U_Z(s))  \mathcal{D}(U_X(t))  \omega_k^Z \ .
\end{equation}
We consider the first part (a row vector)
\begin{equation}
	\mathbf{e}_s = \mathbf{e} \ \mathrm{e}^{  \mathcal{D}(Z)  s} =   (\omega_k^z)^\dagger   \mathcal{D}(X)^\dagger  e^{   \mathcal{D}(Z) s} \ , 
\end{equation}
for which entries are $0$ apart from entries $k \pm 1$:
\begin{equation}
	(\mathbf{e}_s)_{k \pm 1} = i  \frac{\sqrt{j(j+1)}}{2}  \ e^{ \mp   i s} \ .
\end{equation}
We can write this as:

\begin{equation}
\begin{split}
& \mathbf{e}_s =  \\ & (0 , ... ,  i \frac{\sqrt{j(j+1)}}{2}   e^{    i s} , 0 ,  i \frac{\sqrt{j(j+1)}}{2}   e^{ -   i s} , 0 , ... , 0) \ .
\end{split}	
\end{equation}

We now determine the second part of the function $f(s,t)$: $\mathcal{D}(U_X(t)) \omega_k^Z$. Since $\omega_k^Z$ has a single entry equal to 1 at position $k$ this just selects the $k^{th}$ column of $\mathcal{D}(U_X(t))$, we call these states $\omega_X(t)$; they form an orbit. The inner product $\mathbf{e}_s \cdot \omega_X(t)$ will only have two terms corresponding to the $k-1$ and $k+1$ entries of each vector. 
\begin{equation}
\begin{split}
& \mathbf{e}_s \cdot \omega_t = \\
& i \frac{\sqrt{j(j+1)}}{2}    (e^{i s} \omega_X(t)_{k-1} +   e^{ - i s} \omega_X(t)_{k+1}) \ .
\end{split}  
\end{equation}
This is real valued for all values of $s$, hence it must be the case that $\omega_X(t)_{k-1} = \omega_X(t)_{k+1}$ and both are imaginary (or both real and $\omega_X(t)_{k-1} = - \omega_X(t)_{k+1}$).  However since the $\omega_k^Y$ state belongs to the orbit $\omega_X(t)$ and has the property $(\omega_k^Y)_{k-1} = (\omega_k^Y)_{k+1}$ it must be the former.
\begin{equation}
\mathbf{e}_s \cdot \omega_t = \sqrt{j(j+1)} \omega_X(t)_{k+1} \mathrm{cos}(s) \ .
\end{equation}
The maxima and minima of this function occur for $s=0$ (or $\pi$ but this corresponds to the same orbit). 
Hence we need to find the maxima and minima of the function:

\begin{equation}
	\begin{split}
	f(s,t=0) = g(t)  & = 
	\mathbf{e} \cdot \mathcal{D}(U_X  (t))  \omega_k^Z
	\\ &= (\omega_k^Z)^\dagger  \mathcal{D}(X)^\dagger \cdot \mathcal{D}(U_X  (t))  \omega_k^Z
	\\ & = (\omega_k^X)^\dagger  \mathcal{D}(Z)^\dagger \cdot \mathcal{D}(U_Z (t))  \omega_k^X \ .
	\end{split} 
\end{equation}

Where we have used $H$ the Hadamard transformation defined above which changes the basis from $Z$ to $X$: $H Z  H^\dagger = X$. Moreover $H  U_Z(t)  H^\dagger = U_X(t)$. Therefore:
\begin{equation}
\mathcal{D}(U_X(t))  \omega_k^Z   = \mathcal{D}(H) \mathcal{D}(U_Z(t))  \mathcal{D}(H)^\dagger \omega_k^Z \ ,
\end{equation}
we note that $\mathcal{D}(H)^\dagger  \omega_k^Z$ is a 0 eigenstate of $\mathcal{D}(X)$ and is therefore equal to $\omega_k^X $.

We now evaluate the second part of $g(t): \mathcal{D}(U_Z(t)) \omega_k^X$ where $\omega_k^X$ is a normalised zero eigenvector of $\mathcal{D}(X)$ and has the following form:
\begin{equation}
  \omega_k^X 
  = 
  \left(\begin{array}{c}
  a_j  \\ 
  0  \\ 
  a_{(j-2)}  \\
  0 \\
  \vdots \\
  0 \\
  -a_{(j-2)} \\
  0 \\
  -a_j
\end{array} \right) \ ,
\end{equation}
where $|a_n|>|a_m|$ for $n>m$.

\begin{equation}
  	\mathcal{D}(e^{ Z t})  \omega_k^X
  = 
  \left(\begin{array}{c}
  a_j e^{ i   j  t }  \\ 
  0  \\ 
  a_{(j-2)} \ e^{ i  (j-2)  t } \\
  0 \\
  \vdots \\
  0 \\
  -a_{(j-2)} e^{-i  (j+2)  t } \\
  0 \\
  -a_{j} e^{-i    j  t }
\end{array} \right)  \ .
\end{equation}

We can also determine $(\omega_k^X)^\dagger  \mathcal{D}(Z)^\dagger$:
\begin{equation}
\begin{split}
& (\omega_k^X)^\dagger  \mathcal{D}(Z)^\dagger =   \\
& (j a_j^*, 0 , (j-2) a_{j-2}^*, 0 , ...., 0 ,  (j-2) a_{j-2}^* , 0 ,  j a_j^*) \ .
\end{split}
\end{equation}
We can now compute $g(t)$:
\begin{equation}
	\begin{split}
	g(t)  & = (\omega_k^X)^\dagger  \mathcal{D}(Z)^\dagger \cdot \mathcal{D}(U_Z (t))  \omega_k^X
		\\ & = 2  \sum_{l = 1,\ l \ \mathrm{odd}}^j \  l \ |a_l|^2 \ \mathrm{sin}(  kt) \ .
	\end{split} 
\end{equation}
$|a_n|>|a_m|$ for $n>m$ hence $|a_n|^2>|a_m|^2$ for $n>m$. We restrict ourselves to the interval $[0, 2 \pi)$.
We have been considering non-quantum theories with $j>1$, however we briefly describe what this effect would correspond to for the quantum case. When $j=1$ (i.e $k = 2$) there is a single maximum occurring for $t_m = \pi/2$. Hence the global minimum occurs for $t = 3 \pi /2$. The two states distinguished by this effect are antipodal and correspond to the $\mathcal{D}(X)$ eigenstates. Indeed we observe that this effect is just given by the tangent to $\omega_2^Z$ (image of the ray $\ket{0}$) in the $\mathcal{D}(X)$ direction which is proportional to $\omega_2^X$. We know that this effect gives the outcome probability of being in the $\omega_2^X$ state (image of the $\ket{+}$ ray) which is maximal for $\omega_2^X$ and minimal for $-\omega_2^X$ (image of the $\ket{-}$ ray).

We now show that $g(t)$ has two global maxima and two global minima for $j>1$. $g(t) = g(\pi - t)$, therefore given a maximum/minimum we can find another (unless $t_m = \pi/2$). Moreover since $g(t) = - g(t+\pi)$ given a maximum/minimum we can find a minimum/maximum.
To prove our claim we need to find a global maximum/minimum in the interval $[0 , \pi)$. If this extremum occurs for a value which is not $\pi/2$ then we can find the other maximum/minimum in the same interval and the two minima/maxima in $[\pi, 2 \pi)$.
We now show that for $j>1$ the global maximum/minimum does not occur for $t_m = \pi/2$.

We first note that $g(0)=0$, hence a global maximum is positive and a global minimum is negative.
We first compute $g(\pi/2)$ and show that:  $g(\pi/2)>0 \rightarrow g(\pi/(2j))>g(\pi/2)$. This implies that $g(\pi/2)$ cannot be a global maximum, since if it is positive there is a $\tau=\pi/(2j)$ such that $g(\tau) > g(\pi/2)$. Similarly we show that $g(\pi/2)<0 \rightarrow g(\pi/(2j))<g(\pi/2)$ which entails that when $g(\pi/2)$ is negative it cannot be a global minimum. Hence the global extrema of $g(t) $ do not occur for $t= \pi/2$ when $j>1$.
\begin{equation}
	g(\pi/2) = \sum_{l =1 ,l \ \mathrm{odd}}^j (-1)^{\frac{l-1}{2}} l \ |a_l|^2  \ .
\end{equation}
We note that:
\begin{equation}
	l  |a_l|^2 > (l-2) |a_{l-2}|^2, \quad l \  \mathrm{odd}, \ l >1 \ .
\end{equation}
Since each term in $g(\pi/2)$ alternates sign and the absolute value of each term increases for each $l$ the sign of $g(\pi/2)$ is determined by that of its highest term  $(-1)^{\frac{j-1}{2}} l \ |a_j|^2$. This implies $g(\pi/2)>0$ for  $j = 5 , 9, 13 ,...$. We consider these cases. The last term (which is the largest) $(-1)^{\frac{j-1}{2}} j \ |a_j|^2$  is always positive. We observe that the sum of the remaining terms is negative. 
This implies:
\begin{equation}\label{Hjeq}
g(\pi/2) -j |a_j|^2 < 0 \ ,
\end{equation}
We now determine
\begin{equation}
	g(\frac{\pi}{2j}) = \sum_{l =1 , l \ \mathrm{odd}}^j l \ |a_l|^2 \ \mathrm{sin}(\frac{k \pi}{2j}) \ .
\end{equation}
The arguments in each of the $\mathrm{sin}$ functions in $g(\frac{\pi}{2j})$ are always between $0$ and $\frac{\pi}{2}$, hence each term is positive. Therefore:
\begin{equation}\label{Hj2eq}
	g(\frac{\pi}{2j}) - j |a_l|^2 > 0  \ .
\end{equation}
Combining equations \ref{Hjeq} and \ref{Hj2eq} we obtain $g(\frac{\pi}{2j}) > g(\frac{\pi}{2})$ when $g(\frac{\pi}{2})>0$ . This shows that the maximum of the function does not occur for $t= \pi/2$. The function therefore has at least two maxima in the interval $[0,\pi]$. 
	
For $j = 3 , 7 , 11,...$ the same argument can be applied to show that the function has two minima within the interval $[0, \pi]$ and hence two maxima within $[\pi , 2\pi]$.

This entails that there exists a global maximum which is separated from a global minimum by a value which is not $\pi$. Hence there exist states which are distinguishable but are not antipodal.

\subsection{Bit symmetry (Result~\ref{Resultbitsymmetry})}

In theories belonging to $\mathcal{T}_2^I$ images of the orthogonal rays $\psi_0$ and $\psi_1$ are antipodal states $\Omega_{\psi_0}$ and $\Omega_{\psi_1}$ with $\Omega_{\psi_0} = -\Omega_{\psi_1}$. These states can be perfectly distinguished using the measurement $(\mathbf{e},\frac{1}{2}),(-\mathbf{e}^T,\frac{1}{2})$ where $\mathbf{e} = \Omega_{\psi_0}^T/2$. 
From Result $\ref{Rnonantipodal}$ there exists a state $\Omega_{\psi_2}$ which is distinguishable from $\Omega_{\psi_0}$ and not antipodal to it. Due to the faithfulness of $\Omega$ for $j$ odd we have $\psi_2 \neq \psi_1$.
Since $\psi_1$ is the unique ray orthogonal to $\psi_0$ in $\mathrm{P}\mathbb{C}^2$  $\psi_2$ is not orthogonal to $\psi_0$. There is no unitary which maps the pair of orthogonal rays $(\psi_0,\psi_1)$ to the pair of non-orthogonal rays $(\psi_0,\psi_2)$. Hence there exist pairs of distinguishable states which are not related by a reversible transformation belonging to the dynamical group.

\subsection{Phase groups in $\mathcal{T}_2^I$ (Result~\ref{phasegroupres})}
	
We show that the maximal measurement (from the proof of Result~\ref{Rnonantipodal}) which distinguishes two non-antipodal states $\Omega_\psi$ and $\Omega_\phi$ in theories $\mathcal{T}_2^I$ cannot have a phase group. Without loss of generality we choose $\psi$ to be invariant under transformations $U_Z(t)$.  The phase group of a measurement is a subgroup of the transformation group which leaves all outcome probabilities of the measurement unchanged.
This measurement  has exactly two maxima and two minima on the state space (as shown in the proof of Result~\ref{Rnonantipodal}). We call $E$ one of the effects which compose this measurement: 
\begin{align*}
E(\Omega_\psi) &= E(-\Omega_\phi) = 1  \ , \\
E(\Omega_\phi) &= E(-\Omega_\psi) = 0 \ .
\end{align*}
The states $\Omega_\psi$ and $\Omega_\phi$ are the images of non-orthogonal rays on $\mathbb{C}^2$. We now show that there is no subgroup of the transformation group which leaves the outcome probabilities of the 4 states $\Omega_\psi$, $\Omega_\phi$, $-\Omega_\psi$ and $-\Omega_\phi$ invariant (which are the images of $\psi$, $\phi$ , $\psi^\perp $ and $\phi^\perp$). This is sufficient that there is no phase group for this measurement. Such a phase group must either preserve the states, or map them to states which give the same outcome probabilities. For example it must either preserve $\psi$ or map it to $\phi^\perp$.
First consider transformations which map $\Omega_\phi$ to itself. These correspond to the $\mathrm{U}(1)$ subgroup $U_Z(t)$ (or a discrete subgroup of this subgroup). These transformations do not map $\Omega_\phi$ to itself (nor to $-\Omega_\psi$) and hence do not preserve the outcome probabilities of the effect.
The only other possible transformations which preserve the outcome probabilities for these 4 states are of the form:
\begin{align*}
U \psi &= \phi^\perp \ , \\
U \phi &= \psi^\perp \ .
\end{align*}
In order for a phase group to exist $U^\dagger$ must also preserve the outcome probabilities: 
\begin{align*}
U^\dagger \psi &= \phi^\perp \ , \\
U^\dagger \phi &= \psi^\perp \ ,
\end{align*}
entailing that $U^\dagger = U$. The only $U \in \mathrm{PU}(2)$ which has this property is the identity. Therefore the only phase group of this maximal measurement is the trivial one.

\subsection{No-simultaneous encoding (Result~\ref{Rnosim})}
	
All theories in $\mathcal{T}_2^I$  have pairs of non-antipodal states $\omega_0$ and $\omega_1$ which are perfectly distinguishable. The measurement which distinguishes them perfectly, also distinguishes $-\omega_0$ and $-\omega_1$. 
\begin{align}
E_i(\omega_j) &= \delta_{ij} \ ,  \\
E_i(- \omega_j) &= -\delta_{ij} + 1 \ .
\end{align}
A first bit $a$ can be perfectly encoded as follows:
\begin{align*}
	a = 0 \rightarrow \omega_0 \ &\mathrm{or} \ -\omega_1 \ , \\
	a = 1 \rightarrow \omega_1 \ &\mathrm{or} \ -\omega_0 \ .
\end{align*}
The second bit $a'$ can be encoded as:
\begin{align*}
	a' = 0 \rightarrow \omega_0 \ &\mathrm{or} \ \omega_1 \ , \\
	a' = 1 \rightarrow -\omega_1 \ &\mathrm{or} \ -\omega_0 \ .
\end{align*}
For example if Alice needed to encode the bits $a = 0$ , $a' = 0$ she would choose the state $\omega_0$. According to the scenario Alice encodes her bits $a$ and $a'$ in a single system and sends it to Bob. He then tries to guess one of the bits. If he chooses to guess the value of bit $a$ he can do so with certainty (using the measurement which perfectly distinguishes $\omega_0$ and $-\omega_1$ from $-\omega_0$ and $\omega_1$ ). If Bob chooses to guess the value of bit $a'$ he can use another effect to obtain partial information about whether the state is $\omega_0 \ \mathrm{or} \ \omega_1 $ or whether the state is $-\omega_1 \ \mathrm{or} \ -\omega_0$. This could be any effect which partially distinguishes $\frac{\omega_0 + \omega_1}{2}$ from $- \frac{\omega_0 + \omega_1}{2}$. Since neither of these is the maximally mixed state such an effect exists.

\subsection{Pure-state dual theories (Result~\ref{PSDres})}
	
A two outcome measurement is a pair of effects $E_1 = (\mathbf{e},c)$ and $E_2 = (-\mathbf{e},1-c)$. In theories with irreducible $\Gamma$ (for $d=2$) for each pure state $\Omega_\psi \in \mathcal{S}$ (where $\mathcal{S}$ is the state space) the state $-\Omega_\psi$ also exists. For a \textit{pure state dual} theory we impose that effects are proportional to states and tight. Hence the effects are given by: 
\begin{equation}
\{\mathbf{e} = \frac{\Omega^T_\psi}{2} | \psi \in \mathrm{P}\mathbb{C}^2 \} \ .
\end{equation} 
The set of effects is such that for every $\mathbf{e}$ the linear functional $-\mathbf{e}$ also exists. Two outcome measurements are of the form $M = \{(\mathbf{e},\frac{1}{2}),(-\mathbf{e},\frac{1}{2})\}$. The linear functional $\mathbf{e}$ has a single maximum for state $\Omega^T_\psi$ and a single minimum for $-\Omega^T_\psi$. Hence the only states which are distinguishable are antipodal. This entails that the state spaces are bit symmetric. The fact that for all effects there is a single maximum/minimum entails that no-simultaneous encoding holds. Given a functional $\mathbf{e}$ there is a $\mathrm U (1)$ subgroup which leaves $\Omega_\psi$ and hence $\mathbf e $ invariant. Hence every measurement $M =\{(\mathbf{e},\frac{1}{2}),(-\mathbf{e},\frac{1}{2})\}$ has a $\mathrm{U}(1)$ phase group.

\subsection{Bit symmetry in $\mathcal{T}_2^R$ (Result~\ref{Bitsymred})}

Consider a theory in $\mathcal{T}_2^R$; its state space $\mathcal{S}$ is a direct sum of state spaces of theories in $\mathcal{T}_2^I$:
\begin{equation}
\mathcal{S} = \bigoplus_{j \in \mathcal{J}} \mathcal{S}_j \ .
\end{equation} 
Moreover $\mathcal{S}$ has at least one faithful block in the decomposition. In the case where this block is non-quantum we denote it by $k$ ($k>1$, $k$ odd). We assume all effects are allowed. Now consider effects with support solely on subspace $k$. The state space restricted to this subspace is just a state space corresponding to a theory with $j=k$ ( and hence in $\mathcal{T}_2^I$). There exists a measurement (with support just in this subspace) which can distinguish pairs of non-antipodal states (by Result~\ref{Rnonantipodal}). Moreover there also exists a measurement (with support only in this subspace) which distinguishes pairs of antipodal states. Hence the entire state space $\mathcal{S}$ has pairs of distinguishable states which are images of orthogonal rays and pairs which are not images of orthogonal rays.

If the generators contain a single faithful block $j=1$ (with all others unfaithful) then a measurement on that block can distinguish a pair of states (which are images of orthogonal rays). The unfaithful blocks have pairs of distinguishable states which are not images of orthogonal rays. This follows from the fact that orthogonal rays are mapped to the same state in unfaithful representations and that the unfaithful block are also state spaces when considered alone. Since all state spaces have at least two distinguishable states it follows that pairs of states which are not images of orthogonal rays can be distinguished using effects with support in the unfaithful blocks.

Therefore all theories in $\mathcal{T}_2^R$ have pairs of distinguishable states which are images of orthogonal rays and pairs which are not; these theories violate bit symmetry.

\subsection{Proof of Result~\ref{resultgend}}

In order to show that all unrestricted state spaces are not bit symmetric we first need to establish how a $\mathrm{P}\mathbb{C}^{d-1}$ state space is embedded in a $\pcd$ state space. In the following we assume that all state spaces are unrestricted.
\begin{lemma}\label{Lrestrictingtosubspace}
	Given a theory of $\mathrm{P}\mathbb{C}^d$ corresponding to maps $\Omega$ and $\Gamma$, where the highest weight representation in $\Gamma$ is $\mathcal{D}_j^d$, the image of a $\mathrm{P}\mathbb{C}^{d-1}$ subspace of $\pcd$ under $\Omega$ is equivalent to a state space of a $\mathrm{PU}(d-1)$ theory with representation $\Gamma'$, whose highest weight component is $\mathcal{D}_j^{d-1}$. 
\end{lemma}
This lemma is proven in appendix \ref{L3}. It follows from the lemma that any non-quantum $\mathrm{P}\mathbb{C}^3$ (i.e. which has a block $\mathcal{D}_j^3$ with $j>1$) state space when restricted to a $\mathrm{P}\mathbb{C}^2$ subspace is equivalent to a non-quantum state space with a representation which has a block $\mathcal{D}_j^2$ (if $j$ is even it must also be the case that the representation has a block with $j$ odd since the representation of states is faithful). Hence the restricted state space belong to $\mathcal{T}_2^R$. By Result \ref{Bitsymred} this $\mathrm{P}\mathbb{C}^2$ state space is not bit symmetric, that is to say it has pairs of distinguishable states which have different (Hilbert space) inner products. Moreover the $\mathrm{P}\mathbb{C}^3$ state space is not bit symmetric since there are no transformations in $\mathrm{PU}(3)$ mapping pairs of states with different Hilbert space inner products.

Any non-quantum $\pcd$ state space (i.e. which has a block $\mathcal{D}_j^d$ with $j>1$) has a  $\mathrm{P}\mathbb{C}^{d-1}$ subspace which is equivalent to a state space associated to a reducible representation which has a block $\mathcal{D}_j^{d-1}$. If the $\mathrm{P}\mathbb{C}^{d-1}$ subspace is not bit symmetric (i.e. the Hilbert space inner product between two pairs of distinguishable state is not the same) then the $\pcd$ state space is not bit symmetric either (since even considering the whole $\mathrm{PU}(d)$ transformation group its elements will still only map between pairs of states with the same inner product).

Hence by induction any non-quantum $\pcd$ ($d \geq 2$) unrestricted state space is not bit symmetric.

\subsection{Proof of Lemma~\ref{branchingrule}}\label{Lemma1proof}

\subsubsection{Partitions, $\mathrm{U}(d)$ and $\mathrm{SU}(d)$}

A partition $\lambda$ is a set of $d$ integers such that $\lambda_1 \geq \lambda_2 \geq ... \geq \lambda_{d-1} \geq \lambda_{d} \geq 0$. We define the size of the partition $|\lambda| = \sum_{i = 1}^n \lambda_i = k$ and say that $\lambda$ is a partition of $k$ into $d$ parts.  We use square brackets to denote $\lambda = [\lambda_1 , ... , \lambda_d]$.

Each partition $\lambda$ is associated to an irreducible representation $\pi_d^\lambda$ of $\ud$~\cite{Whipman_branching_1965}. Moreover any irreducible representation $\pi_d^\lambda$ of $\ud$ is also an irreducible representation of $\sud$ \cite{Whipman_branching_1965} . By this we mean that if we consider a subgroup $\pi_d^\lambda(U), \ U \in \sud$ this representation is not reducible. Two irreducible representations $\lambda^{(1)}$ and $\lambda^{(2)}$ of $\ud$, when restricted to $\sud$, are equivalent representations of $\sud$ if and only if $\lambda^{(1)}_i - \lambda^{(2)}_i$ is constant for all $i$ \cite{Whipman_branching_1965}. 

A representation of $\sud$ corresponding to a partition $\lambda$ and a representation of $\sud$ corresponding to $\lambda + c$ are therefore equivalent. Hence an irreducible representation of $\sud$ is characterised by the differences between elements of a partition $\lambda$. We can map $\lambda$ to $(\lambda_1 - \lambda_2, \lambda_2 - \lambda_3, ... , \lambda_{d-1} - \lambda_d)$ which is a vector of positive integers of size $d-1$ which is the same for $\lambda $ and $\lambda + c$ for all $c$ . This vector $(\lambda_1 - \lambda_2, \lambda_2 - \lambda_3, ... , \lambda_{d-1} - \lambda_d)$ is the previously introduced Dynkin index (which is in one-to-one correspondence with irreducible representations of $\sud$). We use square brackets to differentiate a partition from a Dynkin index. In the following it will be useful to describe representations of $\sud$ using their partition (even if the map from partitions to representations of $\sud$ is many to one). Since partitions differing by a constant give the same representation we fix $\lambda_d = 0$. This requirement fixes a representation of $\sud$ for each $\lambda$. 

\subsubsection{$\mathrm U (1) \times \mathrm{SU} (d-1)$ invariant subspaces}

For a representation $\lambda$ of $\mathrm{SU}(d)$ acting on $V$ to have an $\mathrm U (1) \times \mathrm{SU} (d-1)$-invariant vector $v$ means the following. If we take the representation of a $\mathrm U (1) \times \mathrm{SU} (d-1)$ subgroup (which is reducible) and consider its action on $v$, it leaves $v$ invariant. The reducible representation of $\mathrm U (1) \times \mathrm{SU} (d-1)$ is composed of irreducible blocks. If we decompose $v$ according to these blocks, we see that the components of $v$ in these subspaces (for non-trivial representations of $\mathrm U (1) \times \mathrm{SU} (d-1)$) will not be left invariant by all of the transformations $\mathrm U (1) \times \mathrm{SU} (d-1)$ (since each block is irreducible, so cannot have invariant subspaces within it). However if one of the blocks corresponds to a trivial representation of $\mathrm U (1) \times \mathrm{SU} (d-1)$ then it will leave that component of $v$ invariant. Thus we can choose $v$ to have support just in those subspaces which correspond to trivial representations in the decomposition of $\lambda$ when restricting to a $\mathrm U (1) \times \mathrm{SU} (d-1)$ subgroup.

We now find the representations $\lambda$ such that there exists at least one trivial representation of $\mathrm U (1) \times \mathrm{SU} (d-1)$ when $\lambda$ is restricted to this $\mathrm{U}(1) \times \mathrm{SU}(d-1)$ subgroup. 

\subsubsection{Branching rule for $\ud \rightarrow \mathrm{U}(d-1)$}

Restricting representations of groups to subgroups leads to a reducible representation of the subgroup given by a branching rule. We introduce the necessary tools.

We consider a partition $\lambda = [\lambda_1 , ... , \lambda_n]$. A partition $\mu = [\mu_1 , ... , \mu_{n-1}]$  is said to interlace $\lambda$ when:
\begin{equation}
\lambda_1 \geq \mu_1 \geq ... \geq \lambda_{n-1} \geq \mu_{n-1} \geq \lambda_n \ 
\end{equation}
The branching rule from $\ud$ to $\mathrm{U}(d-1)$ is as follows. Let $\mathrm{U}(d)$ have irreducible representation $\pi_d^\lambda$ acting on a space $V_\lambda$. Then there is a unique decomposition of $V_\lambda$ into subspaces under the action of $\mathrm{U}(d-1)$ \cite[p.19]{Holman1971}:
\begin{equation}
V_\lambda = \bigoplus_\mu V_\mu \ ,
\end{equation}
where the sum is over every $\mu$ which interlaces $\lambda$ and  $V_\mu$ is a carrier space for an irreducible representation of $\mathrm{U}(d-1)$ labelled by $\mu$.

\subsubsection{Branching rule for $\sud \rightarrow \mathrm{SU}(d-1) \times \mathrm{U}(1)$}

The restriction of $\pi_d^\lambda$ (a representation of $\mathrm{U}(d)$) to an $\mathrm{SU}(d)$ subgroup is an irreducible representation of $\mathrm{SU}(d)$ with partition $\lambda$ (which for $\mathrm{SU}(d)$ is defined up to a constant). This representation acts irreducibly on $V_\lambda$. The subgroup $\mathrm{U}(d-1)$ is given by:
\begin{equation}
\left(\begin{array}{c|c}  
1 & 0\ \ \ \cdots\ \ \ 0 
\\ \hline
0 & \\
\vdots & U_{(d-1) \times (d-1)} \\
0 &    
\end{array}\right), \ U_{(d-1)}U_{(d-1)}^\dagger = \unity \ .
\end{equation}
It's action on $V_\lambda$ decomposes as a direct sum of $V_\mu$. We wish to consider the action of $\mathrm{SU}(d-1)$ on $V_\lambda$. This subgroup is given by:
\begin{equation}
\left(\begin{array}{c|c}  
1 & 0\ \ \ \cdots\ \ \ 0 
\\ \hline
0 & \\
\vdots & U_{(d-1) \times (d-1)} \\
0 &    
\end{array}\right), \  U_{(d-1) \times (d-1)}  \in \mathrm{SU}(d-1) \ .
\end{equation}
$\mathrm{U}(d-1)$ acts on each $V_\mu$ irreducibly and restricting an irreducible representation of $\mathrm{U}(d-1)$ to  $\mathrm{SU}(d-1)$  gives an irreducible representation with the same partition. Hence the $\mathrm{SU}(d-1)$ acts irreducibly on each  $V_\mu$ in the decomposition. The branching rule $\sud \rightarrow \mathrm{SU}(d-1)$ is the same as $\mathrm{U}(d) \rightarrow \mathrm{U}(d-1)$~\cite{Whipman_branching_1965}. 

We now establish the branching  $ \ud \rightarrow \mathrm{SU}(d-1) \times \mathrm{U}(1)$ subgroup. The $\mathrm{U}(1)$ part corresponds to all matrices $C$ of the following form:
\begin{equation}
C = \left(\begin{array}{cccc}  
e^{it} & \ \ \ &\ \  & \\ 
& e^{\frac{-i}{d-1}t} &  & \\
& & \ddots & \\
& &    &  e^{\frac{-i}{d-1}t}
\end{array}\right) \ .
\end{equation}
We note that this $\mathrm{U}(1)$ subgroup commutes with the $\mathrm{SU}(d-1)$ subgroup hence its action will leave the subspaces $V_\mu$ invariant. Therefore the space $V_\lambda$ decomposes into $V_\mu$ invariant subspaces under the action of $\mathrm{SU}(d-1) \times \mathrm{U}(1)$.
We now establish the action of the $\mathrm{U}(1)$ subgroup on these subspaces.
We use a similar technique as that found in proof of theorem 8.1.2 of~\cite[p.364]{Goodman09}. We consider a matrix $A \in \ud$:
\begin{equation}
A = \left(\begin{array}{cccc}  
e^{it} & \ \ \ &\ \  & \\ 
 & e^{it} &  & \\
 & & \ddots & \\
 & &    &  e^{it}
\end{array}\right) \ ,
\end{equation}
And a matrix $B \in \mathrm{U}(d-1)$:
\begin{equation}
B = \left(\begin{array}{cccc}  
1 & \ \ \ &\ \  & \\ 
& e^{-i \frac{d}{d-1}t} &  & \\
& & \ddots & \\
& &    &  e^{-i \frac{d}{d-1}t}
\end{array}\right) \ .
\end{equation}
We note that $A B = C$ for the above defined $\mathrm{U}(1)$ subgroup. We determine the action of $A$ and $B$ on $V_\lambda$ which will allow us to know that of $C$. The action of $A$ on the whole carrier space $V_\lambda$ is multiplication by a scalar $e^{it|\lambda|}$. Similarly the action of $B$ (which belongs to the $\mathrm{U}(d-1)$ subgroup and commutes with all elements of $\mathrm{U}(d-1)$) on each subspace $V_\mu$ is multiplication by $e^{-i \frac{d}{d-1}t |\mu|}$~\cite[theorem 8.1.2. (proof) , p.364]{Goodman09}. Hence we can compute the action of the $\mathrm{U}(1)$ subgroup on each subspace which is multiplication by :
\begin{equation}
e^{it|\lambda|}e^{-i \frac{d}{d-1}t |\mu|} = e^{i t (|\lambda| - \frac{d}{d-1} |\mu|)} \ ,
\end{equation}
Hence the action of this $\mathrm{U}(1)$ subgroup on the carrier space $V_\lambda$ acts by scalar multiplication on each $V_\mu$ (where the scalar can be the same for different $\mu$). 
We can now summarise: given a representation $\pi_d^\lambda$ of $\mathrm{SU}(d)$ acting on $V_\lambda$ there exists a decomposition (into invariant subspaces) under the action of $\mathrm{SU}(d-1) \times \mathrm{U}(1)$ given by:
\begin{equation}
V_\lambda = \bigoplus_\mu V_\mu \ ,
\end{equation}
where the sum is over every $\mu$ which intertwines $\lambda$. The $\mu$ determine irreducible representations of $\mathrm{SU}(d-1)$ acting on each subspace. The $\mathrm{U}(1)$ parts acts like $e^{i t (|\lambda| - \frac{d}{d-1} |\mu|)}$ on each subspace.

\subsubsection{Representation of $\sud$ with trivial representations under the action of $\mathrm{SU}(d-1) \times \mathrm{U}(1)$}

We can now address the problem of finding which representations of $\sud$ are such that their restriction to $\mathrm{SU}(d-1) \times \mathrm{U}(1)$ contains a trivial representation of $\mathrm{SU}(d-1) \times \mathrm{U}(1)$. 

A trivial representation of $\mathrm{SU}(d-1)$ has Dynkin coefficients $(0,...0)$ and hence partition $\mu = [\mu_1 , ... \mu_{d-1}]$ where $\mu_1 = \mu_2 = ... = \mu_{d-1}$. Hence representations of $\sud$ which contain trivial representations of $\mathrm{SU}(d-1)$ in this decomposition will have partitions $\lambda$ where $\lambda_2 = \lambda_3 = ... = \lambda_{n-1} = \mu_1$, following the requirement that $\mu$ intertwine $\lambda$.

We now consider the requirement that the $\mathrm{U}(1)$ action is trivial. Since the action of $\mathrm{U}(1)$ on $V_\mu$ is given by  $ e^{i t (|\lambda| - \frac{d}{d-1} |\mu|)}$, it is  trivial when $|\lambda| - \frac{d}{d-1} |\mu| = 0$. We note that some authors multiply the $\mathrm U (1)$ charge by a constant so that it is an integer value. Since we are considering only 0 $\mathrm U (1)$ charge this will not concern us.

We can now add this requirement on $\lambda$ to the preceding one: $\lambda_2 = \lambda_3 = ... = \lambda_{d-1} = \mu_1$. As stated above we are considering $\lambda_d =0$ in order to identify one representation of $\mathrm{SU}(d-1)$ with each partition $\lambda$. We have
\begin{equation}
|\mu| = (d-1) \mu_1 \ ,
\end{equation}
and
\begin{equation}
|\lambda| = \lambda_1 + (d-2) \mu_1 \ ,
\end{equation}
We substitute this into the requirement $|\lambda| - \frac{d}{d-1} |\mu| = 0$:
\begin{equation}
\begin{split}
\lambda_1 + (d-2) \mu_1 - d \mu_1 = & \  0 \\
 \lambda_1 =  &  \ 2 \mu_1
\end{split}
\end{equation}
Hence the representation $\pi_n^\lambda$ of $\sud$ with partition $[2 \mu_1 , \mu_1 , ... , \mu_1  , 0]$ which corresponds to Dynkin label $(\mu_1  , 0 ,...  , 0 , \mu_1)$ meets the requirements. This shows that any representation of $\sud$ with Dynkin label $(j , 0 , ... , 0 , j)$ (for any positive integer $j$) has a trivial subspace under the action of $\mathrm{SU}(d) \times \mathrm{U}(1)$.  We now show that these representations have a single trivial representation in the decomposition. These are representations with partition $\lambda = [2 j, j , ... , j , 0]$. Under the branching rule into $\mathrm{SU}(d-1) \times \mathrm{U}(1)$ trivial representations correspond to partitions $\mu$ have components $\mu_i$ which are all equal (in order for the $\mathrm{SU}(d-1)$ component to be trivial) and these must be equal to $j$. There is a single such $\mu$ of this form.
This shows that there is a unique trivial representation in the decomposition.

The representations of $\sud$ of the form $(j, 0 , ... , 0 , j)$ correspond to the representations $\mathcal{D}_j^d$. These are the only representations of $\sud $ (and $\pud$) which leave a vector invariant under $\mathrm{SU}(d) \times \mathrm{U}(1)$ (moreover this vector is unique up to normalisation).

\subsubsection{A comment on real and complex irreducibility}\label{ReCoirrep}

In this work we are interested in classifying certain representations of the dynamical group $\sud$. These representations act on state spaces which span real vector spaces. 

All the representation theory results used above concern representations acting on complex vectors spaces. A representation which acts on a complex vector space is real if it can be expressed in a basis where all the matrix elements are real. This is the case of the irreducible representations $\mathcal D_j^d$. These representations are irreducible when acting on both complex and real vector spaces. However not all real irreducible representations correspond to complex irreducible representations. 
There are reducible representations acting on a complex vector space whose action is irreducible when acting on a real vector space. We have found all real irreducible representations which are also complex irreducible which meet our criteria. However it may be the case that there are representations which are real irreducible but complex reducible which meet our criteria. If this is the case our approach would not have found them. We now show that there are no such representations satisfying the invariance properties we require.

A real representation $\Gamma$ which is irreducible on a real vector space but reducible on a complex one can be block diagonalised in the following form $\Gamma = \rho \oplus \bar \rho$~\cite[Exercise 	3.39, p. 41]{Fulton91}. For example consider the following representation of $\mathrm{SO}(2) \cong \mathrm{U}(1)$:
\begin{equation}
\left( \begin{array}{cc}
\mathrm{cos}(\theta) & \mathrm{sin}(\theta)  \\ 
-\mathrm{sin}(\theta) & \mathrm{cos}(\theta) 
\end{array}  \right) \ .
\end{equation}
This is irreducible when acting on $\mathbb{R}^2$. However its action on $\mathbb{C}^2$ can be diagonalised as follows:
\begin{equation}
\left( \begin{array}{cc}
\mathrm{e}^{ i\theta} & 0 \\ 
0 & \mathrm{e}^{ -i\theta}
\end{array}  \right) \ .
\end{equation}
Hence there may be representations which are real irreducible which have $\mathrm{SU}(d-1) \times \mathrm{U}(1)$ invariant vectors but are complex reducible. All real irreducible representations are either: complex irreducible or complex reducible of the form $\rho + \bar \rho$.

Let us consider a real irreducible representation $\Gamma$ which is complex reducible. There exists a transformation $L$ such that $\Gamma = L (\rho \oplus \bar \rho) L^{-1}$. We now ask the same question: does there exist a vector which is invariant under the subgroup $H = \mathrm{SU}(d-1) \times \mathrm{U}(1)$. Let us call this vector $v$. We have:
\begin{equation}
\Gamma_{|H} v =v  \ .
\end{equation}
We observe that if such a vector exists then the (complex) vector $L^{-1} v$ is invariant under $L^{-1} \Gamma_{|H} L = \rho_{|H} \oplus \bar \rho_{|H}$. Moreover if there exists a vector $w$ which is left invariant by  $\rho_{|H} \oplus \bar \rho_{|H}$ then the vector $L w$ is left invariant under $\Gamma_{|H}$. Hence if we show that there are no vectors left invariant under $\rho_{|H} \oplus \bar \rho_{|H}$ then there are no vectors left invariant under $\Gamma_{|H}$.

There exists a vector $w$ invariant under $\rho_{|H} \oplus \bar \rho_{|H}$  if and only if this representation (obtained by restricting $\rho \oplus \bar  \rho$)  has at least one trivial component. This is only possible if at least one of the representations $\rho_{|H}$ or $\bar \rho_{|H}$ has one or more trivial components. 

Here $\rho$ and $\bar \rho$ are complex irreducible representations of $\sud$. The only such representations with a  $\mathrm{SU}(d-1) \times \mathrm{U}(1)$ invariant vector are of the form $\mathcal D_j^d$ which is real (and we observe $\mathcal D_j^d = \bar {\mathcal D}_j^d$). Hence there are no representations which are complex reducible but real irreducible which have these invariant vectors.

\subsection{Proof of Lemma~\ref{subgrouplemma}}\label{Lemma2proof}

We consider two $\mathrm{SU}(d-1) \times \mathrm{U}(1)$ stabilizer subgroups of two distinct rays on $\pcd$ ($d \geq 3$).   Consider a ray $\psi$ with stabilizer group $G_\psi$.The Lie algebra $\mathfrak{g}_\psi$ which generates this group has elements $X_\psi$ (corresponding to $\mathrm{U}(1)$)

\begin{equation}
X_\psi = \left(\begin{array}{c|c}  
i  & 0\ \ \ \cdots\ \ \ 0 
\\ \hline
0 & \\
\vdots & -\frac{i }{d-1} \ \mathbb{I}_{d-1} \\
0 &    
\end{array}\right),
\end{equation}
and elements $Y_\psi$ (corresponding to the $\mathrm{SU}(d-1)$ group)
\begin{equation}
Y_\psi = \left(\begin{array}{c|c}  
0 & 0\ \ \ \cdots\ \ \ 0 
\\ \hline
0 & \\
\vdots & A \\
0 &    
\end{array}\right), \ A = -A^\dagger, \ \mathrm{Tr}(A) = 0 \ .
\end{equation}
As can be seen by the generators $\psi$ is the unique ray stabilized by this subgroup. Hence distinct rays $\psi$ and $\psi'$ have distinct stabilizer groups (which are equivalent up to conjugation). This is not the case for $\mathrm{SU}(2)$ for example; the $\mathrm{U}(1)$ stabilizer of $\ket{0}$ also stabilizes $\ket{1}$.

The two subgroups $G_\psi$ and $G_{\psi'}$ (which stabilizes $\psi'$) are \textit{maximal} subgroups of $\sud$~\cite{antoneli_maximal_2006}. A maximal subgroup $H$ of $G$ is a proper subgroup (i.e. $H\neq G$) such that if $H \leq K \leq G$ then $H= K$ or $K =G $.  The group $H$ generated by these two groups $G_\psi$ and $G_{\psi'}$ is not equal to either $G_\psi$ and $G_{\psi'}$. Hence it is equal to the full group $\sud$.

\subsection{Proof of Lemma~\ref{Lrestrictingtosubspace}}\label{L3}

\subsubsection{Embeddedness of simple theories}

In this section we look at theories which do not obey the constraint given by (\ref{span O}) and study how $\mathrm{P}\mathbb{C}^{d-1}$ subspaces are embedded in $\pcd$ state space. We consider a $\pcd$ state space, which can be generated by applying all transformations $\Gamma_U$ to a reference state $\Omega_0 =  |0 \rangle\!\langle 0 |^{\otimes N}$:
\begin{equation}
\begin{split}
\Gamma_U \Omega_0 &= U^{\otimes N} |0 \rangle\!\langle 0 |^{\otimes N}\, U^{\dagger \otimes N}\ \\ &= \mathrm{Sym}^N U |0 \rangle\!\langle 0 |^{\otimes N}\, \mathrm{Sym}^N U^\dagger ,
\end{split}
\end{equation}
The action of the product $\mathrm{Sym}^N U \otimes \mathrm{Sym}^N U^\dagger$ can be decomposed (using known rules for Young tableaux) as: 
\begin{equation}
\begin{split}
& (N,0,...,0) \otimes (0,...,0,N)  \\
& = (0,...,0) \oplus  (1,0,...,0,1) \oplus ... \oplus (N,0,...,0,N) \ .
\end{split} 
\end{equation}
In our previous notation this is just:
\begin{equation}\label{simpledecomposition}
\Gamma_U = \bigoplus_{i=0}^N \mathcal{D}_i^d \ .
\end{equation}
The state space is given by Hermitian matrices acting on the symmetric subspace of $(\mathbb{C}^d)^{\otimes n}$. However it is not immediately clear that every summand in (\ref{simpledecomposition}) acts on the space of Hermitian matrices (which is the state space). For example when we generate the state space by acting with a reducible representation on a reference state, if the reference state does not have support in every block of the representation then there are certain components of the representation which are superfluous (since they do not act on the state space). The dimension of the space of Hermitian matrices is:
\begin{equation}
D_\omega =   \binom {d+n-1} {n} ^2
\end{equation}
Using the dimension formula for the representations $\mathcal{D}_j^d$ given by (\ref{dimension}) and requiring that the representation of the transformations acts on a space of dimension $D_\omega$  we see that every summand in (\ref{simpledecomposition}) acts on the state space.
Hence we see that these theories are equivalent to reducible theories of $\pcd$ with $\mathcal{J} = \{1,2,...,N \}$. We consider a $\mathrm{P} \mathbb{C}^{d-1} $ subspace of the Hilbert space and see which subspace of the state space it corresponds to. It can be generated from the state $ |1 \rangle\!\langle 1 |^{\otimes N}$ by applying an $\mathrm{SU}(d-1)$ subgroup:
\begin{equation}
U^{\otimes N} |1 \rangle\!\langle 1 |^{\otimes N} U^{\dagger \otimes N}  \ ,
\end{equation}
With:
\begin{equation}
U = 
\left(\begin{array}{c|c}  
1 & 0\ \ \ \cdots\ \ \ 0 
\\ \hline
0 & \\
\vdots & U_{(d-1) \times (d-1)} \\
0 &    
\end{array}\right)
\end{equation}
We notice that this state space is just a $\mathrm{P}\mathbb{C}^{d-1}$ state space of the type described above with representation:
\begin{equation}
\Gamma' = \bigoplus_{i=0}^N \mathcal{D}_i^{d-1} \ .
\end{equation}

\subsubsection{Embeddedness of irreducible theories}

We consider a $\pcd$ theory with representation $\mathcal{D}_j^d$. We want to know the state space corresponding to a $\mathrm{P}\mathbb{C}^{d-1}$ subspace. We consider a basis $\{\ket{0}, \ket{1} , ... , \ket{d-1} \}$ and determine the image of all states of the form $\alpha_1 \ket{1} + ... + \alpha_{d-1} \ket{d-1}$. These are all the states (apart from $\ket{0}$) which are invariant under the  $\mathrm{U}(1)$ group of the $\mathrm{SU}(d-1) \times \mathrm{U}(1)$ subgroup specified in the proof of Lemma~\ref{subgrouplemma}. From Lemma~\ref{U1charge}  (proven in appendix \ref{L4}) we know that restricting the representation $\mathcal{D}_j^d$ to a $\mathrm{SU}(d-1) \times \mathrm{U}(1)$ subgroup gives a reducible representation of $\mathrm{SU}(d-1) \times \mathrm{U}(1)$. Moreover the $\mathrm{U}(1)$ action is trivial for the following blocks:
\begin{equation}
\bigoplus_{i=0}^j \mathcal{D}_i^{d-1} \ .
\end{equation}
The image of $\alpha_1 \ket{1} + ... + \alpha_{d-1} \ket{d-1}$ therefore lies in this subspace (since it is invariant under this $\mathrm{U}(1)$ action) which we call the $\alpha = 0$ subspace. Moreover the image of $\ket{0}$ is uniquely determined as the trivial subspace $\mathcal{D}_0^{d-1}$ (which is invariant under the whole subgroup). However we do not know if the image of the  $\mathrm{P}\mathbb{C}^{d-1}$ subspace spans the whole $\alpha = 0$ subspace. We show that it must have support in the subspace $\mathcal{D}_j^{d-1}$ (which is part of the $\alpha=0$ subspace).

We first consider a reducible theory described with representation:
\begin{equation}
\Gamma = \bigoplus_{i=0}^j \mathcal{D}_i^{d} \ .
\end{equation}
This corresponds to a reducible theory as characterised above. We call its state space $\mathcal{S}$. Considering a $\mathrm{P}\mathbb{C}^{d-1}$ subspace gives a state space $\mathcal{S}'$ with representation:
\begin{equation}
\Gamma' = \bigoplus_{i=0}^j \mathcal{D}_i^{d-1}  \ .
\end{equation}
We observe that the state space with representation $\Gamma$ is equivalent to a direct sum of state spaces with representation $\mathcal{D}_k^{d}$ for $k = 1,...,j$. We label these state spaces $\mathcal{S}_k$. We write:
\begin{equation}\label{redstatespace} 
\mathcal{S} = \bigoplus_{k=0}^j \mathcal{S}_k \ .
\end{equation}
Moreover reducing a state space $\mathcal{S}_k$ gives a $\mathrm{P}\mathbb{C}^{d-1}$ state space with a representation inside the direct sum (i.e. which may or may not have support in each representation): 
\begin{equation}
\bigoplus_{i=0}^k \mathcal{D}_i^{d-1} \ .
\end{equation}
We have seen that the restriction of $\mathcal{S}$ gives a state space $\mathcal{S}'$ with support in block $\mathcal{D}_j^{d-1}$. Moreover since $\mathcal{S}$ is a direct sum of state spaces $\mathcal{S}_k$ it must be the case that (at least) one of these $\mathcal{S}_k$ reduces to a state space with support in $\mathcal{D}_j^{d-1}$. The state space $\mathcal{S}_j$ is the only state space in the direct sum in (\ref{redstatespace}) which can have support in  $\mathcal{D}_j^{d-1}$ when restricted. That is to say the reduction of $\mathcal{S}_j$ (with representation  $\mathcal{D}_j^{d}$) gives a $\mathrm{P}\mathbb{C}^{d-1}$ state space with support in block $\mathcal{D}_j^{d-1}$.

\subsubsection{Embeddedness of arbitrary reducible theories}

From the above considerations we see that an irreducible theory $\mathcal{J} = j$ of $\mathrm{P} \mathbb{C}^{d}$ when restricted to a $\mathrm{P}\mathbb{C}^{d-1}$ subspace must give a state space which has support in the subspace $\mathcal{D}_j^{d-1}$, and hence corresponds to a theory of $\mathrm{P}\mathbb{C}^{d-1}$ which has a representation which contains at least a block $\mathcal{D}_j^{d-1}$. A reducible theory of $\pcd$ $\mathcal{J} = j_1 , ..., j_n$ has a state space $\mathcal{S}$ which is a direct sum of state spaces $\mathcal{S}_{j_i}$. The restriction of $\mathcal{S}$ to a  $\mathrm{P}\mathbb{C}^{d-1}$ subspace is equivalent to restricting each of the $\mathcal{S}_{j_i}$ state spaces. Hence the restriction of  $\mathcal{S}$ to $\mathrm{P}\mathbb{C}^{d-1}$ will have support in at least subspaces $\mathcal{D}_k^{d-1}$ for $k = j_1 , ... ,j_n$. 

The Lemma follows directly from this.

\subsection{Proof of Lemma~\ref{U1charge}}\label{L4}

\begin{lemma}\label{U1charge}
Consider a representation $\mathcal{D}_j^d$ and its decomposition under a $\mathrm{SU}(d-1) \times \mathrm{U}(1)$ subgroup. The reducible representation of $\mathrm{SU}(d-1) \times \mathrm{U}(1)$ has blocks with various $\mathrm{U}(1)$ charges. The subspace with $0$ $\mathrm{U}(1)$ charge is acted upon by the representation of $\mathrm{SU}(d-1)$:
\begin{equation}
\bigoplus_{i=0}^j \mathcal{D}_{d-1}^i \ .
\end{equation}
\end{lemma}
\noindent Similarly to Lemma~\ref{branchingrule} we label $\mathcal{D}_j^d$ with partition $\lambda = [2 j , j , j .... , j , 0]$. The restriction to a $\mathrm{SU}(d-1) \times \mathrm{U}(1)$ subgroup acts reducibly on the carrier space $V_\lambda$ as:
\begin{equation}
V_\lambda = \bigoplus_\mu V_\mu \ ,
\end{equation}
where the sum is over every $\mu$ which intertwines $\lambda$ and  $V_\mu$ is a carrier space for an irreducible representation of $\mathrm{SU}(d-1)$ with partition $\mu$ . The $\mu$ which intertwine $\lambda$ are of the form:
\begin{equation}
\mu = [a , j , j , ... , b] , \ 2j \geq a \geq j \ , j \geq b \geq 0 \ .
\end{equation}
We have $|\lambda| = d j$ and $| \mu| = (d-3) j +a+b$. From Lemma \ref{branchingrule} we have the condition that the $\mathrm{U}(1)$ charge is 0 when $|\lambda| - \frac{d}{d-1} |\mu| = 0$. We now substitute in the relevant expressions:
\begin{equation}
\begin{split}
& |\lambda| - \frac{d}{d-1} |\mu| = 0 \\
 &  d j -  \frac{d}{d-1} ((d-3) j +a+b) = 0 \\
 &  (d-1) j - (d-3) j - a  - b = 0 \\
& a = 2j - b \ .
\end{split}
\end{equation}
This entails that when $b = 0$, $a = 2j$, when $b = 1 , a = 2j -1$... and when $b= j$, $a = j$.   Every possible $\mu$ (intertwining $\lambda$) is in the direct sum and hence every $\mu$ which meets the above condition for having 0 $\mathrm{U}(1)$ charge is in the decomposition. Hence there are $j+1$ subspaces $V_\mu$ where the action of $\mathrm{U}(1)$ is trivial. If we express the representations acting on these in terms of Dynkin notation we observe that $\mu = [2j-b , j , ... , j , 2j-b]$ becomes $(j-b , 0 , 0 ... , j-b)$ for $b = {0,...,j}$. The terms in the direct sum are therefore just the representations $\mathcal{D}_i^{d-1}$ for $ i = {0, ... , j}$.

\end{document}